\documentclass[aps,prd, floatfix,notitlepage,nofootinbib,twocolumn,superscriptaddress]{revtex4-1}
\usepackage{amsmath}
\usepackage{amssymb}
\usepackage{graphicx}
\usepackage[tight]{subfigure}
\usepackage{enumitem}
\usepackage{soul}
\usepackage{cancel}
\usepackage{tikz}
\usepackage{pifont}
\usepackage{xspace}
\usepackage{comment}

\makeatletter
\renewcommand{\p@subsection}{}
\renewcommand{\p@subsubsection}{}
\makeatother

\usepackage[english]{babel}
\makeatletter
\def\bbl@set@language#1{%
  \edef\languagename{%
    \ifnum\escapechar=\expandafter`\string#1\@empty
    \else\string#1\@empty\fi}%
  \@ifundefined{babel@language@alias@\languagename}{}{%
    \edef\languagename{\@nameuse{babel@language@alias@\languagename}}%
  }%
  \select@language{\languagename}%
  \expandafter\ifx\csname date\languagename\endcsname\relax\else
    \if@filesw
      \protected@write\@auxout{}{\string\select@language{\languagename}}%
      \bbl@for\bbl@tempa\BabelContentsFiles{%
        \addtocontents{\bbl@tempa}{\xstring\select@language{\languagename}}}%
      \bbl@usehooks{write}{}%
    \fi
  \fi}
\newcommand{\DeclareLanguageAlias}[2]{%
  \global\@namedef{babel@language@alias@#1}{#2}%
}
\makeatother
\DeclareLanguageAlias{en}{english}

\usetikzlibrary{arrows}
\usetikzlibrary{intersections}
\usetikzlibrary{shapes.geometric}
\usetikzlibrary{decorations.pathmorphing, patterns,shapes,fixedpointarithmetic}
\usetikzlibrary{decorations.markings}

\tikzset{
  mid arrow/.style={postaction={decorate,decoration={
        markings,
        mark=at position .575 with {\arrow{stealth}}
      }}},
  end arrow/.style={postaction={decorate,decoration={
        markings,
        mark=at position 1 with {\arrow{stealth}}
      }}},
  snake arrow/.style={fixed point arithmetic, decorate, decoration={snake,amplitude=2pt, segment length=11pt},postaction={decoration={markings,mark=at position 0.625 with {\arrow{stealth}}},decorate}},
}

\usepackage{bm}

\newcommand{\ket}[1]{|#1 \rangle}
\newcommand{\bra}[1]{\langle #1|}
\usepackage[pdfusetitle,colorlinks=true,citecolor=blue,linkcolor=magenta]{hyperref}

\graphicspath{{./}{./images/}}
\begin{document}
\title{Measurement-induced phase transitions in quantum automaton circuits}

\author{Jason Iaconis}
\affiliation{Department of Physics and Center for Theory of Quantum Matter, University of Colorado, Boulder CO 80309, USA}

\author{Andrew Lucas}
\affiliation{Department of Physics and Center for Theory of Quantum Matter, University of Colorado, Boulder CO 80309, USA}

\author{Xiao Chen}
\affiliation{Department of Physics, Boston College, Chestnut Hill, MA 02467, USA}

\begin{abstract}
    We study the entanglement dynamics in a generic quantum automaton circuit subjected to projective measurements. We design an efficient algorithm which not only allows us to perform large scale simulation for the R\'enyi entropy but also provides a physical picture for the entanglement dynamics, which can be interpreted in terms of a classical bit-string model which belongs to the directed percolation universality class. We study the purification dynamics of a state formed by EPR pairs, and the growth of entanglement starting from a product state. In both cases, we verify numerically that the dynamics is in the universality class of classical directed percolation.
\end{abstract}

\maketitle

\section{Introduction}
Recent years have seen a surge of interest in understanding if and how quantum systems subjected to random measurement can nevertheless protect quantum information.  A particularly illuminating example of how this can happen has been discovered by studying random unitary circuits subjected to random projective measurements.  In these systems, there is a phase transition from a volume law entanglement phase at low measurement rate $p<p_c$, to an area law entanglement phase at high measurement rate $p>p_c$ \cite{Li_2018,Chan_2019,Skinner_2019,Li_2019,gullans2019purification,gullans2019scalable,Tang_Zhu_2020,jian2019measurementinduced,Bao_2020,Choi_2020,Szyniszewski_2019,goto2020measurementinduced,Cao_Tilloy_2019,fan2020selforganized,vijay2020measurementdriven,lavasani2020measurementinduced,sang2020measurement,ippoliti2020entanglement,fidkowski2020dynamical,Turkeshi_2020}.  It has been proposed that the volume law phase may be relevant for the construction of quantum error correction schemes for near term devices \cite{Choi_2020,gullans2019purification}.


The literature largely focuses on two classes of hybrid quantum circuits that mix unitary dynamics and projective measurements.   Random Haar circuits \cite{Nahum_2017,Nahum_2018,von_Keyserlingk_2018} have been used to obtain an analytical understanding of this measurement-induced transition \cite{Skinner_2019,Chan_2019,Bao_2020,jian2019measurementinduced}: in particular, the phase transition of the Hartley entropy can be understood by finding a minimal cut on a lattice with broken bonds \cite{Nahum_2017}.
However, numerical simulation of these systems is difficult, and is limited to system sizes of $\le$30 qubits \cite{Skinner_2019,zabalo2019critical}.  Random Clifford circuits \cite{Nahum_2017,Chan_2019,Li_2018,Li_2019,li2020conformal,gullans2019purification,gullans2019scalable,lavasani2020measurementinduced,sang2020measurement,Choi_2020,li2020statistical} allow for large scale numerical simulations, through which accurate critical exponents associated with the transition can be obtained.  Unfortunately, as of now, there is almost no analytic understanding of the measurement-induced phase transition in a random Clifford circuit.

In this paper, we present a new class of hybrid quantum circuit, which both allows for large scale numerical simulations, and also gives a simple analytic understanding of the phase transition in terms of classical directed percolation (DP) \cite{hinirchsen2008non}. In our model, the unitary dynamics is governed by a quantum automaton (QA) circuit, which has been successfully used to simulate unitary quantum dynamics with unusual symmetries up to very large system sizes \cite{Gopalakrishnan2018,Gopalakrishnan_Zakirov2018,PhysRevB.100.214301,iaconis2020quantum,Chen_2020,chen_gu_lucas}, using efficient classical MC simulations.  The special feature of QA circuits is that they take product states (in the computational Pauli $Z$ basis, e.g.) into product states.  Nevertheless, QA circuits are generally able to create highly entangled wave functions. 
By projectively measuring single spins in the Pauli $Z$ basis, coupled by a subsequent rotation of the measured spin into the $X$ basis, we find non-trivial dynamics which also exhibits a measurement-induced phase transition from volume law to area law entanglement, as measured by the second R\'enyi entropy. 


We describe the resulting entanglement phase transition in terms of a classical bit-string model, which belongs to the DP universality class \cite{hinirchsen2008non}. As a consequence, the QA circuit qualitatively differs from the previously studied random Haar circuit and random Clifford circuit, each of which has emergent two dimensional conformal symmetry at the critical point \cite{Skinner_2019,gullans2019purification,li2020conformal}. As DP is anisotropic between space and time, the phase transition is not described by a conformal field theory; instead, it has dynamical critical exponent $z=1.581$ \cite{hinirchsen2008non}. We perform large scale simulation in various hybrid QA circuits. We investigate a Clifford QA circuit in which the unitary dynamics is constructed from a subset of Clifford gates. We compute the entanglement entropy in this model in terms of the stabilizer language \cite{Aaronson_2004,gottesman1998heisenberg}. We also consider a more general non-Clifford QA circuit model and compute the second R\'enyi entropy by using the MC algorithm. In both cases, we verify that they belong to the DP universality class with $z=1.581$.

\section{Review of QA model and algorithm}
\subsection{Review of QA Model} \label{sec:QAreview}

Throughout this work, we consider dynamics generated by quantum automaton (QA) gates. This is a unitary evolution of the quantum wave function which does not generate entanglement when applied to product states in a specific basis (which we refer to as the computational basis).  When acted on by an automaton gate $U$, a computational basis state $\ket{m}$ generally evolves as
\begin{eqnarray}
U \ket{m} = e^{i \theta_m} \ket{\pi(m)}
\label{eq:QA_gate}
\end{eqnarray}
where $\pi(m)$ is an element of the permutation group  on the $2^N$ basis states $\ket{m}$. In this paper, we choose $\ket{m}$ to be a product state in the Pauli $Z$ basis: $m$ is a bit-string which can take 0 or 1 at each site.

\begin{figure}[t]
 \includegraphics[width=.7\columnwidth]{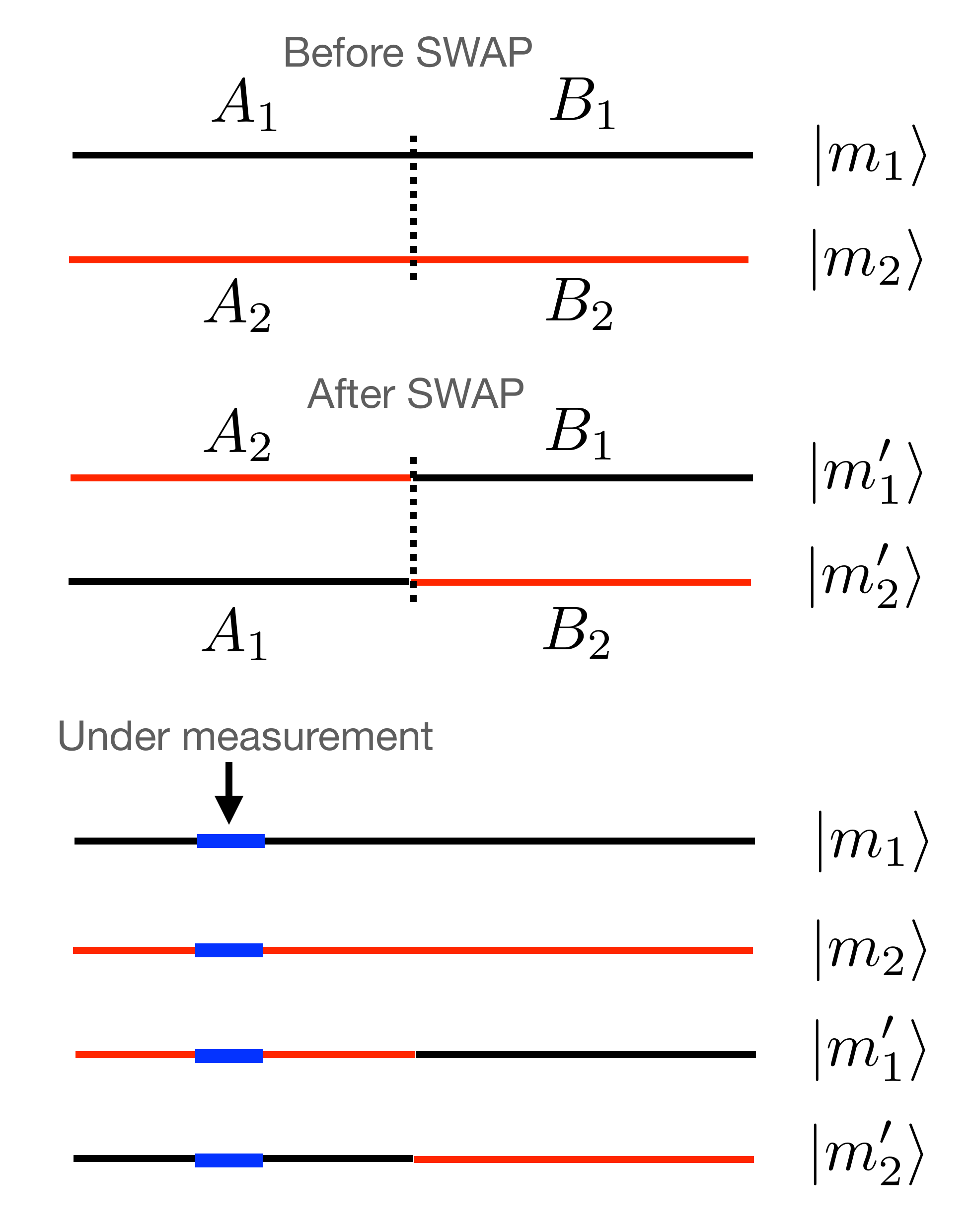}
\caption{A cartoon for the $\mathsf{SWAP}$ operation and the composite measurement operation $M_{i,\sigma}$.} 
\label{fig:SWAP}
\end{figure}

When $U$ acts on an initial state which is a product state in a basis perpendicular to the computational basis, it will generically create entanglement.  Specifically, when acting on the state $\ket{\psi_0}$, which is polarized in the x-direction
\begin{eqnarray}
\ket{\psi_0} \, = \, \bigotimes_i \ket{+x} 
           \,  =  \,\frac{1}{\sqrt{2^N}} \sum_m \ket{m},
           \label{eq:x_polarized}
\end{eqnarray}
we have that
\begin{eqnarray}
U \ket{\psi_0} = \frac{1}{\sqrt{2^N}} \sum_m e^{i\theta_m} \ket{\pi(m)}.
\end{eqnarray}
The presence of the random relative phases $e^{i \theta_m}$ will generically lead to highly complex volume law entangled wave functions $\ket{\psi(t)}$. As explained in \cite{PhysRevB.100.214301, iaconis2020quantum}, such an evolution can be simulated using probabilistic ``variational Monte Carlo" (MC) methods.  Specifically, we can estimate expectation values of different operators $\mathcal{O}$, by sampling the evolution of only a small subset of states  from the full exponential Hilbert space.  For example, we can evaluate the expression
\begin{eqnarray}
\langle \mathcal{O} \rangle &=& \bra{\psi_0} U^\dagger \mathcal{O} U \ket{\psi_0} \nonumber\\
&=& \frac{1}{2^N}\sum_{m,n} e^{i(\theta_m-\theta_n)} \bra{\pi(n)} \mathcal{O}  \ket{\pi(m)}.
\end{eqnarray}
If $\mathcal{O}$ is a simple automaton operator, i.e., $\mathcal{O}\ket{m_1}=\ket{m_2}$, we can estimate this expectation value by performing the full forward and backward time evolution on the sampled states $\ket{m}$. We use the fact that $U^\dagger \mathcal{O}U \ket{m} = e^{i\theta_m} \ket{m^\prime}$, so that

\begin{eqnarray}
   \langle \mathcal{O} \rangle &\approx& \frac{1}{M} \sum_{m=1}^M \bra{m^\prime} U^\dagger \mathcal{O} U\ket{m} \nonumber\\
   &=& \frac{1}{M} \sum_{m=1}^M e^{i\theta_m},
   \label{eq:O_exp}
\end{eqnarray}
where $M$ is the sample number.
Importantly, we can use this method to calculate the entanglement of the state $\ket{\psi(t)}= U \ket{\psi_0}$. This is because for the $n^{\text{th}}$ R\'enyi entropy,
\begin{eqnarray}
S_A^{(n)} &=& \frac{1}{1-n} \log_2(\mbox{Tr}[\rho_A^n]) \nonumber\\
\rho_A &=& \mbox{Tr}_B \ket{\psi}\bra{\psi},
\end{eqnarray}
$\mbox{Tr}[\rho_A^n]$ is equivalent to the $\mathsf{SWAP}$ operator expectation defined on the $n$ copies of the state \cite{PhysRevLett.104.157201}. 
Specifically, in this work, we will focus on the second R\'enyi entropy $S_A^{(2)}$, which can be estimated by measuring the $\mathsf{SWAP}$ operator in a ``doubled'' geometry using
\begin{eqnarray}
    \mbox{Tr}\rho_A^2 = \bra{\psi_0}_1\bra{\psi_0}_2 (U^\dagger \otimes U^\dagger) \, \mathsf{SWAP} \, (U\otimes U) \ket{\psi_0}_1\ket{\psi_0}_2, \nonumber \\
\end{eqnarray}
where $\ket{\psi_0}_i$ are identical copies of the wave function defined in Eq.~\ref{eq:x_polarized}, which exist on separate `layers' of the doubled geometry.
We can partition the entangled wave function $U \ket{\psi_0}$ into two regions $A$ and $B$
\begin{eqnarray}
\ket{\psi} = U \ket{\psi_0} = \frac{1}{\sqrt{2^N}}\sum_{i,j} e^{i\theta_{ij}} \ket{\alpha_i}_A \ket{\beta_j}_B.
\end{eqnarray} 
The $\mathsf{SWAP}$ operator swaps the spin configurations in the A region between the two layers of the full wave function (See Fig.~\ref{fig:SWAP}). We can use Eq.\eqref{eq:O_exp} to efficiently simulate entanglement dynamics in the unitary QA circuit.



\subsection{QA Circuits with Measurements}
\label{sec:QA_meas_algorithm}
We can build on the above formalism to include non-unitary projection operations in QA circuits. Through this, we aim to study the quantum dynamics in hybrid quantum circuits which can show a measurement-induced entanglement phase transition.

Notably, in the classical QA simulation we can not efficiently apply projection operators which project into a subspace which is not the computational basis. Projection into the computational basis is allowed, however, such operations permanently `remove' basis states from our wave function and effectively reduce the rank of the Hilbert space. If the measurement rate $p$ is finite, starting from $\ket{\psi_0}$, after a finite amount of time, it will evolve into a computational basis state -- there is no entanglement in the steady state.

We overcome these issues by introducing a new type of non-unitary dynamics into our circuit, whereby we apply a combination projective measurement and adaptive rotation to single qubits \cite{vijay2020measurementdriven}.  We refer this as the composite measurement.
Specifically, we apply a single qubit $Z$ measurement, immediately followed by a  randomly chosen $\pm \frac{\pi}{2}$ rotation in the $y$ direction (Hadamard gate, up to orientation), so that the measured spin becomes either $\ket{\pm x}$ with equal probability. Therefore in the computational basis, the wave function is again an equal weight superposition over all basis states. 


These non-unitary transformations can also be simulated efficiently. To do this, we must make some key adjustments to our Monte Carlo (MC) algorithm.

Let $M_{i\sigma} = H_iP_{i\sigma}$ be the combination of projective measurement on site i (which projects into the state $\sigma={0,1}$), plus a Hadamard transformation.  Therefore, we have (after normalization)
\begin{eqnarray}
M_{i 0} \ket{0}_i &=& \ket{0}_i + \ket{1}_i \nonumber \\
M_{i 0} \ket{1}_i &=& 0.
\end{eqnarray}
If we apply a hybrid circuit $\widetilde{U}$, which contains some finite density of $M_{i\sigma}$ operations, we find that for almost all basis states $\ket{m}$, $\widetilde{U} \ket{m} = 0$. Therefore the approach introduced in Eq.\eqref{eq:O_exp} is not useful here. One would need to sample an exponential number of states $\ket{m}$, before finding one in which the configuration at each step matches all corresponding $M_\sigma$ operators. 

Instead, we notice that after  $M_{i\sigma}$ acts on the state $\ket{\psi(t)}$, due to the rotation back to the $x$ basis, we are once again left with a superposition over all computational basis states.  Therefore, $M_{i\sigma} \ket{\psi}$ will always have a nonzero overlap with any basis state $\bra{n}$. Specifically,
\begin{align}
    \bra{n}M_{i\sigma}\ket{\psi}=\bra{T_{i\sigma}(n)} \psi\rangle= \frac{e^{i\theta_{T_{i\sigma}(n)}}}{\sqrt{2^N}}
\end{align}
where $\ket{T_{i\sigma}(n)}$ is equivalent to the state $\ket{n}$ with the spin at site $i$ forced to be in the $\sigma$ state.


This leads to a very simple algorithm for simulating these hybrid quantum circuits with non-unitary dynamics.  For any operator $\mathcal{O}$, we can evaluate its expectation value as
\begin{eqnarray}
\langle \mathcal{O} \rangle &=& \sum_m \bra{\psi_0}\widetilde{U}^\dagger \mathcal{O} \ket{m} \bra{m} \widetilde{U} \ket{\psi_0} . \label{eq:nonunitary_op}
\end{eqnarray}
where
\begin{eqnarray}
\bra{m} \widetilde{U} \ket{\psi_0} \, &=& \quad\,\,\, \bra{m} M_t U_t M_{t-1}U_{t-1}\dots M_1U_1 \ket{\psi_0} \nonumber\\
\, &=&\quad\,\,\, \bra{T_t(m)} U_t M_{t-1} U_{t-1} \dots M_1 U_1 \ket{\psi_0} \nonumber \\
&=&\quad\ \ \dots
\end{eqnarray}
That is, we can evaluate the circuit evolution ``inside out'', by inserting a complete set of states $\ket{m}\bra{m}$ and evaluating the evolution $\bra{m}\widetilde{U}\ket{\psi_0}$ left to right. When we encounter a composite measurement operator, $M_{i\sigma}$, it simply acts on the state $\bra{m}$ by forcing the spin at site $i$ to be in the $\sigma$ state.  Throughout this evolution, we accumulate a phase $e^{i\theta_m}$, from both the right and left expressions in the sum in Eq.~\ref{eq:nonunitary_op}. Therefore,
\begin{eqnarray}
\langle \mathcal{O} \rangle &=& \frac{1}{2^N}\sum_m e^{i \Theta_{1,m}} e^{i\Theta_{2,m}} \nonumber\\
e^{i \Theta_{1,m}} &=& \sqrt{2^N}\bra{\psi_0} \widetilde{U}^\dagger \mathcal{O} \ket{m} \nonumber\\
e^{i \Theta_{2,m}} &=& \sqrt{2^N}\bra{m} \widetilde{U} \ket{\psi_0}.
\end{eqnarray}

Importantly, we can apply this non-unitary algorithm to evaluate the second R\'enyi entropy. Consider again the expression
\begin{eqnarray}
    &&  \mbox{Tr}\rho_A^2=\bra{\psi_0}_1\bra{\psi_0}_2(\widetilde{U}^\dagger \otimes \widetilde{U}^\dagger)\mathsf{ SWAP} (\widetilde{U}\otimes \widetilde{U})\ket{\psi_0}_1\ket{\psi_0}_2 \nonumber\\
   &&  =\sum_{m_1,m_2}\bra{\psi_0}_1\bra{\psi_0}_2(\widetilde{U}^\dagger \otimes \widetilde{U}^\dagger)\mathsf{ SWAP}|m_1\rangle|m_2\rangle \nonumber \\
    &&\hspace{30mm} \times \langle m_1|\langle m_2|(\widetilde{U}\otimes \widetilde{U})\ket{\psi_0}_1\ket{\psi_0}_2 \nonumber\\
    && = \sum_{m_1,m_2}\bra{\psi_0}_1\bra{\psi_0}_2(\widetilde{U}^\dagger \otimes \widetilde{U}^\dagger)|m_1^\prime\rangle|m_2^\prime\rangle \nonumber \\
    &&\hspace{30mm} \times \langle m_1|\langle m_2|(\widetilde{U}\otimes \widetilde{U})\ket{\psi_0}_1\ket{\psi_0}_2 \nonumber\\
    &&\hspace{8.5mm}=\frac{1}{4^N}\sum_{m_1,m_2} e^{i\Theta_{1,m}}e^{i\Theta_{2,m}},
    \label{eq:EE}
\end{eqnarray}
where
\begin{align}
    e^{i\Theta_1}&\equiv 2^N\bra{\psi_0}_1\bra{\psi_0}_2(\widetilde{U}^\dagger \otimes \widetilde{U}^\dagger)\mathsf{ SWAP}|m_1\rangle|m_2\rangle\nonumber\\
    &=2^N\bra{\psi_0}_1\widetilde{U}^\dag\ket{m_1^\prime}\bra{\psi_0}_2\widetilde{U}^\dag\ket{m_2^\prime}\nonumber\\
     e^{i\Theta_2}&\equiv 2^N\langle m_1|\langle m_2|(\widetilde{U}\otimes \widetilde{U})\ket{\psi_0}_1\ket{\psi_0}_2\nonumber\\
     &=2^N\bra{m_1}\widetilde{U}\ket{\psi_0}_1\bra{m_2}\widetilde{U}\ket{\psi_0}_2.
    \label{eq:theta_phase}
\end{align} 
In the hybrid circuit, we evaluate the same unitary on the four states $\{\ket{m_1},\ket{m_2},\ket{m_1^\prime},\ket{m_2^\prime}\}$, where $\ket{m_1^\prime}\ket{m_2^\prime} = \mathsf{SWAP} \ket{m_1}\ket{m_2}$, as shown in Fig.~\ref{fig:SWAP}. The action of $M_{i,\sigma}$ is then to force the spin configuration at site $i$ to be equal in all four states. 

When the spin configurations are equal, any random phases generated in Eq.~\ref{eq:EE} are cancelled, i.e., $\Theta_{1,m}=-\Theta_{2,m}$. This leads to a very physical picture of entanglement growth in hybrid QA circuits. For the two bit-strings $A_1B_1$ and $A_1B_2$, initially, they are differed only in the B region. Under a general unitary QA evolution, this difference will spread to region A, and eventually spread over the entire system.  On the other hand, the non-unitary measurements will force spins on the same site to be equal.  Therefore, there are two opposing forces which attempt to force the two bit-strings to either diverge to completely different configurations or converge to a common configuration. We will show in the next section that this classical bit-string model belongs to the DP universality class and the competition between these two forces is responsible for the entanglement transition in the hybrid QA circuit.


\section{Purification transition and the connection with DP universality class}
In Ref.~\onlinecite{gullans2019purification}, Gullans and Huse explored the purification transition in a hybrid random Clifford circuit. The basic idea is to consider system A and environment B entangled together, and subsequently apply the hybrid quantum dynamics consisting of unitary and measurement gates solely on system A.  How does the entanglement between A and B change as a function of time?  While the system A must eventually be purified due to the measurement, there is a phase transition which is captured by the time to purification. When the measurement rate $p<p_c$, the purification time diverges exponentially in the system size. When $p>p_c$, it takes $O(\log L)$ time to purify the system.\footnote{This is because $S_A^{(2)}\sim L\exp(-a t)$, so the time it takes before $S_A^{(2)} < 1$ is $O(\log L)$.} In this section, we consider a similar setup and study the dynamics of purification in our hybrid QA circuit, where we will see a simple interpretation of the purification dynamics in terms of a classical bit-string model which belongs to the DP universality class.

\begin{figure}[hbt]
\centering
\subfigure[]{\label{fig:Clifford_purification_1} \includegraphics[width=.7\columnwidth]{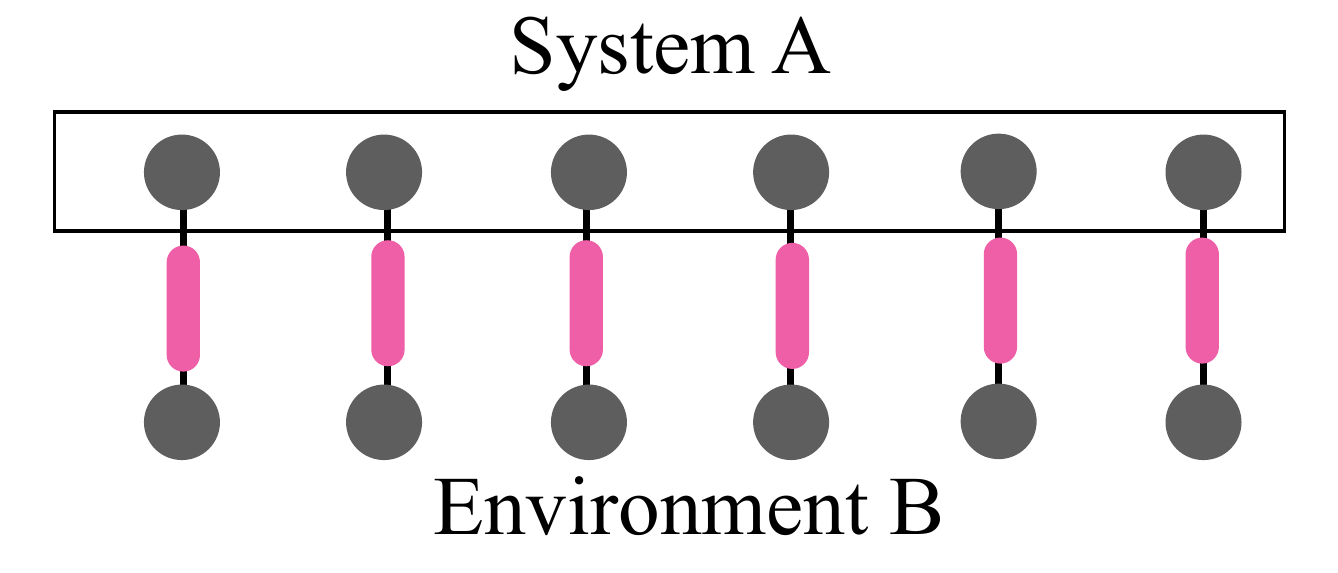}}
\subfigure[]{\label{fig:Clifford_purification_2} \includegraphics[width=.7\columnwidth]{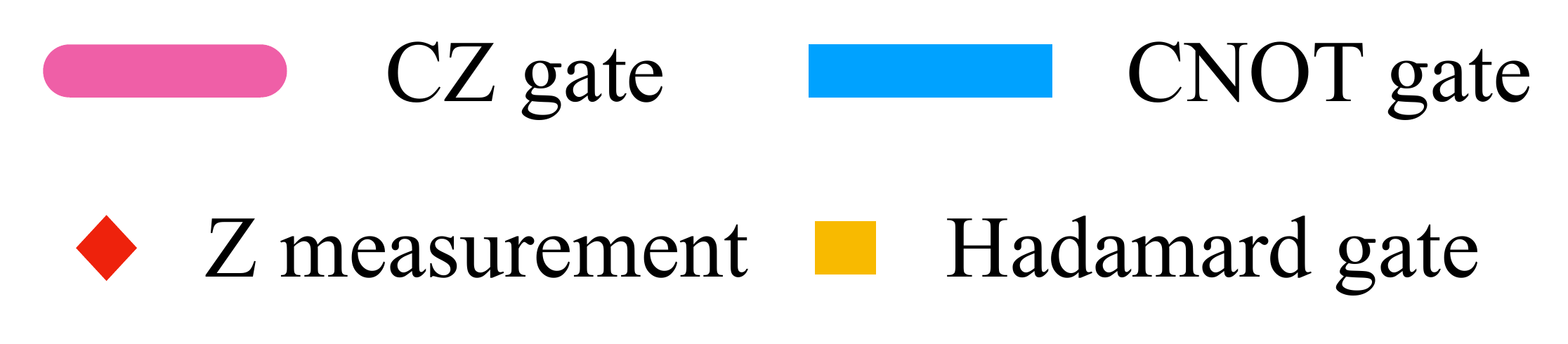}}
\subfigure[]{\label{fig:Clifford_purification_3} \includegraphics[width=.7\columnwidth]{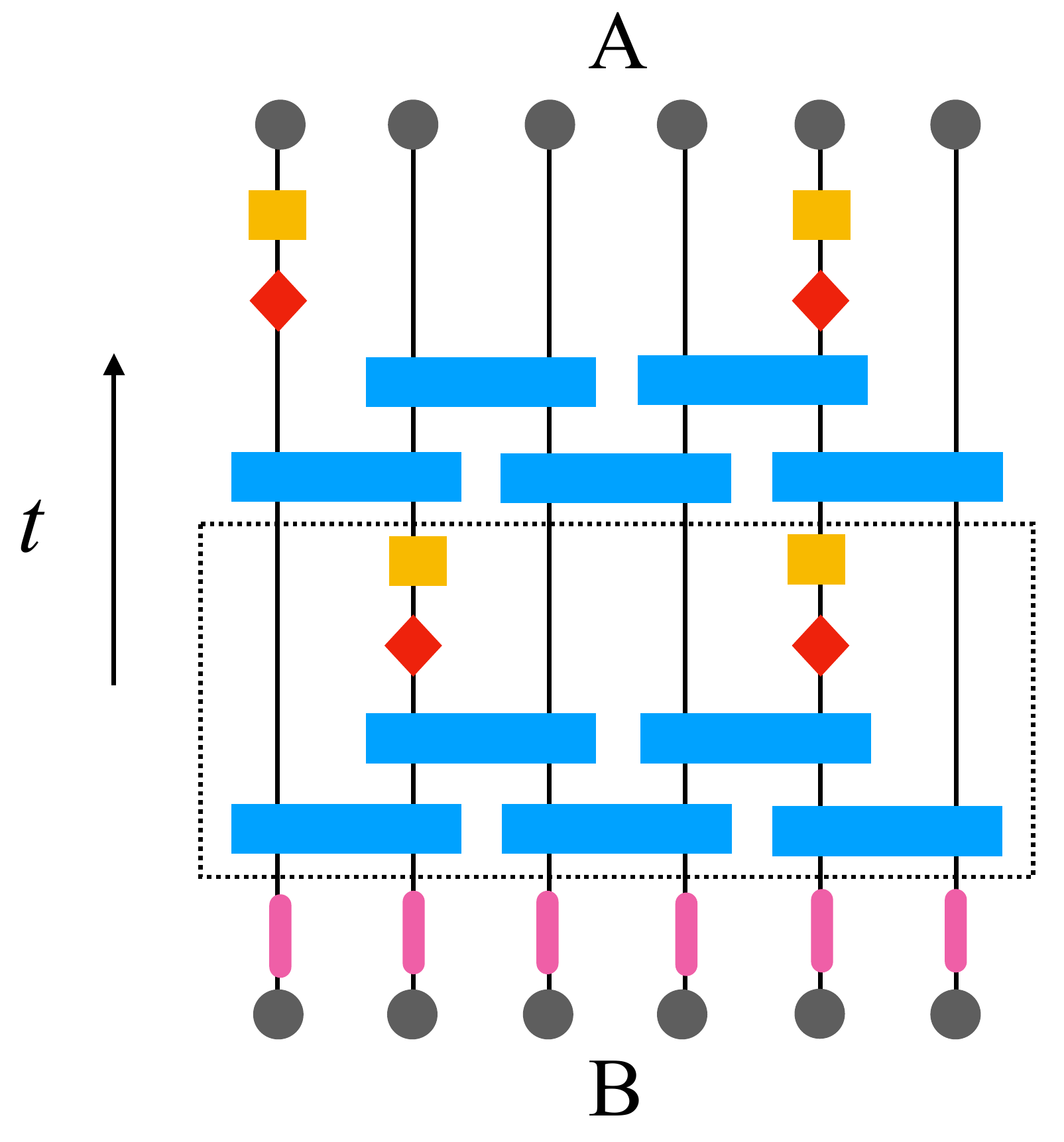}}
\caption{(a) We use CZ gates to form EPR pairs between system A and environment B. (b) We list the single-qubit and two-qubit gates used in circuit (c). (c) We take the initial state in (a) and  apply the hybrid quantum dynamics solely in system A. We study the entanglement between A and B as a function of time. The dashed box denotes time evolution in one time step. Notice that there are two types of CNOT gates. In the first type, the first qubit is the control gate and in the second type, the second qubit is the control gate. We apply them randomly with equal probability. } 
\label{fig:puri_cartoon}
\end{figure}

We take the product state $|\psi_0\rangle$ defined in Eq. \eqref{eq:x_polarized} with $2L$ number of qubits and separate them into system A and environment B with an equal number of qubits.  As shown in Fig.~\ref{fig:Clifford_purification_1}, after applying CZ gates, we form $L$ EPR pairs and the entanglement entropy between A and B is $S_A^{(2)}=L$. Notice that the CZ gate is a two-qubit gate and is defined as follows:
\begin{align}
    \text{CZ}=\begin{pmatrix}
    1 & 0 & 0 & 0\\
    0 & 1 & 0 & 0\\
    0 & 0 & 1 & 0\\
    0 & 0 & 0 & -1\\
    \end{pmatrix}
\end{align}
on the basis $\ket{00},\ket{01},\ket{10},\ket{11}$. It assigns a minus sign for the state $|11\rangle$. 

In terms of Eq.\eqref{eq:EE}, the entanglement entropy $S_A^{(2)}=L$ can be interpreted in this way: for any pair of states $\ket{m_1}$ and $\ket{m_2}$, if they are the same in regime $A$, they are invariant under the $\mathsf{SWAP}$ operator. Therefore $\Theta_1$ and $\Theta_2$ caused by the CZ gate will exactly cancel and each pair will contribute $1/4^{2L}$ to $\mbox{Tr}\rho_A^2$. There are in total $2^{L}\times 2^{2L}$ such pairs. For any other pairs which are differed in regime A, their total contribution is zero. For instance, consider the simplest example when $\ket{m_1}$ and $\ket{m_2}$ are differed at one site in A, i.e.,$A_1=1$ and $A_2=0$ at this site. In B, at the same site, there are in total four possibilities with $B_1B_2=11,10,01,00$. This will give rise to $\exp(i\Theta_1+i\Theta_2)=1,-1,-1,1$ respectively and their total contribution is zero. This idea can be generalized to any pair of $\ket{m_1}$ and $\ket{m_2}$ which are different in system A, and their total contribution to $\mbox{Tr}\rho_A^2$ is zero. Therefore we have
\begin{align}
S^{(2)}_A=-\log_2 \frac{2^L\times 2^{2L}}{4^{2L}}=L.  
\end{align}

\begin{figure*}[t]
\centering
\subfigure[]{\label{fig:puri_transition} \includegraphics[width=.9\columnwidth]{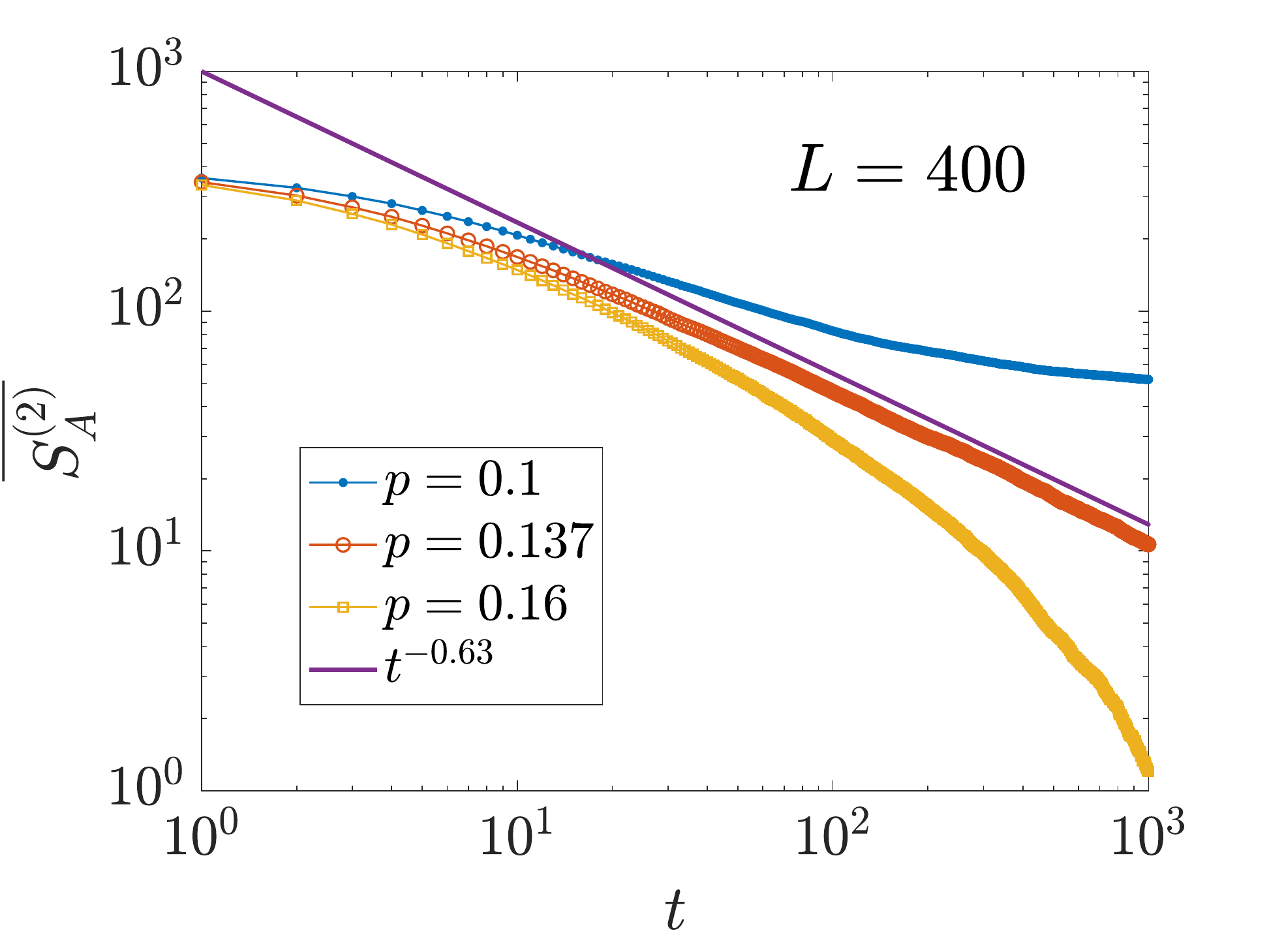}}
\subfigure[]{\label{fig:puri_comparision} \includegraphics[width=.9\columnwidth]{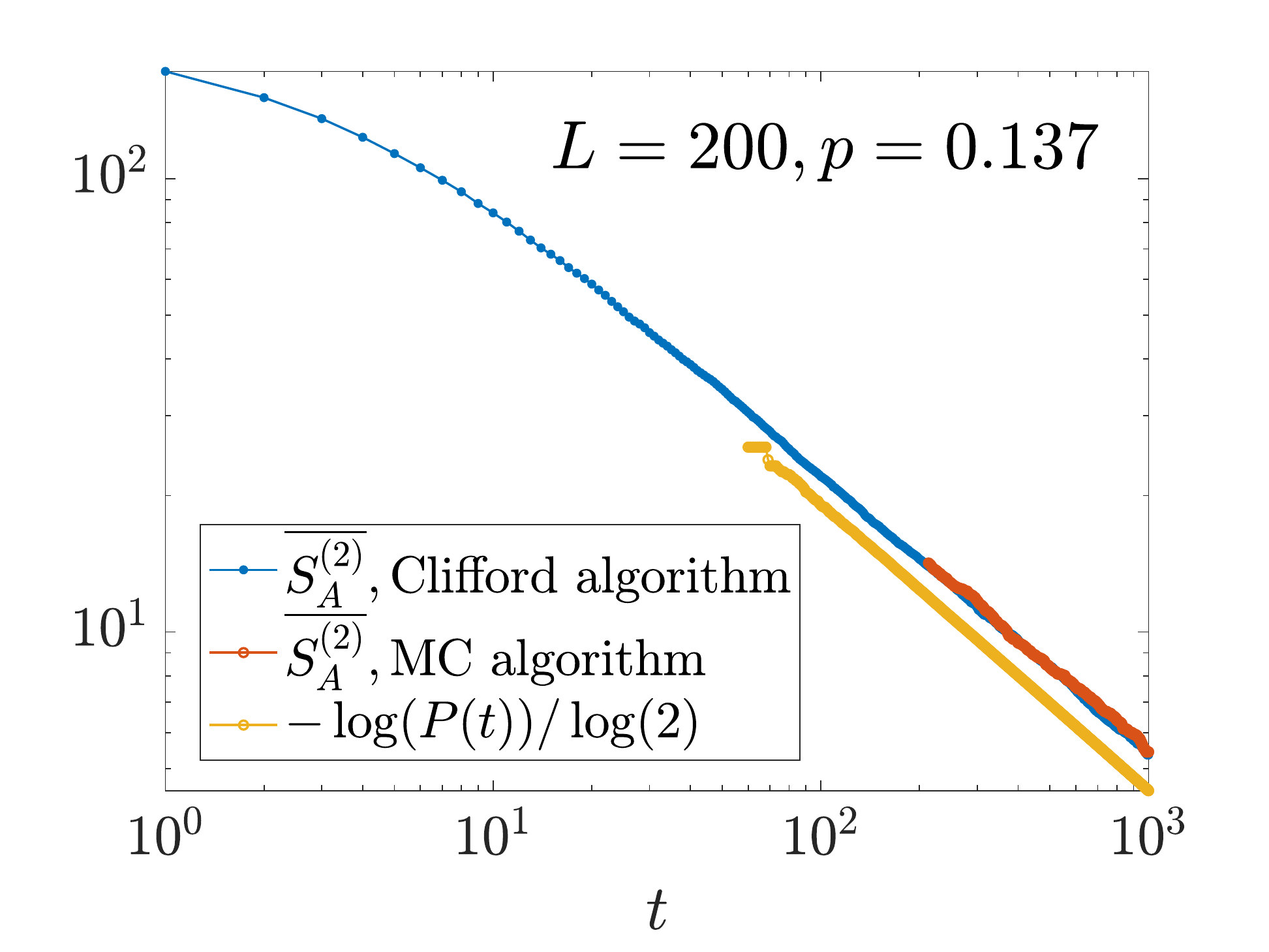}}
\subfigure[]{\label{fig:puri_collapse} \includegraphics[width=.9\columnwidth]{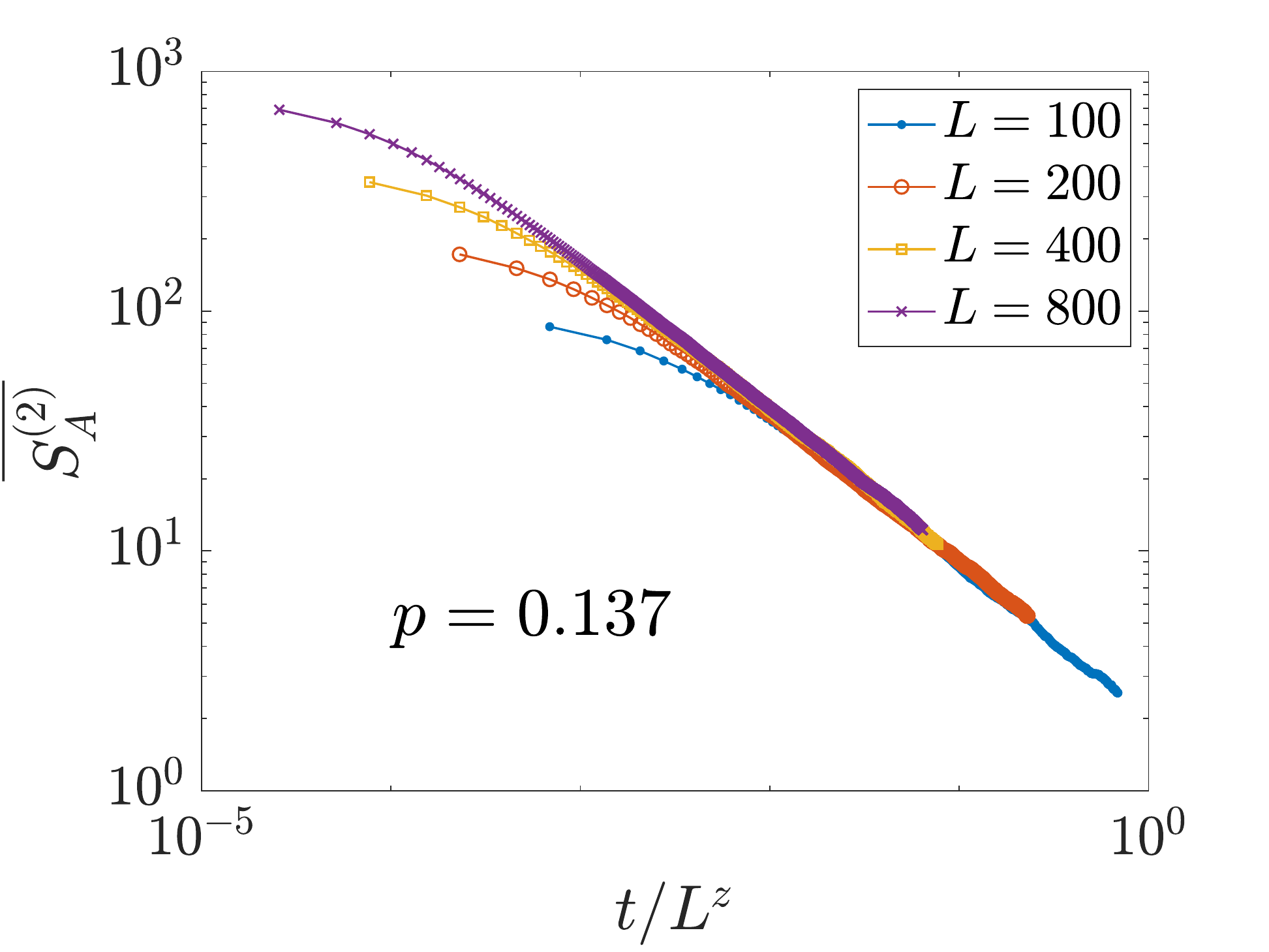}}
\caption{The purification dynamics of $\overline{S_A^{(2)}}$ of the circuit model defined in Fig.~\ref{fig:Clifford_purification_3}. (a) We compute $\overline{S_A^{(2)}}$ at different $p$. At the critical point $p_c=0.137$, $\overline{S_A^{(2)}}$ decays as a power law function with exponent close to $0.63$. (b) We compute $\overline{S_A^{(2)}}$ at the critical point using various methods. One technical issue with the MC algorithm is that $\mbox{Tr}\rho_A^2$ cannot be too small. In the simulation, we manage to compute $\overline{S_A^{(2)}}$ when it is smaller 20. In the last yellow curve, $P(t)$ denotes the probability for a pair of bit-strings to become the same under stochastic time evolution. In this simulation, we take a different stochastic time evolution for different pairs of bit-strings. (c) The data collapse of $\overline{S_A^{(2)}}$ vs the scaling parameter $t/L^z$. In all of these simulations, we take periodic boundary conditions.} 
\label{fig:puri_result}
\end{figure*}
If we introduce unitary gates in system A, the entanglement between A and B remains invariant. As shown in Eq.\eqref{eq:QA_gate}, in QA circuits, the unitary gates will reshuffle the basis and may introduce phases for each $\ket{m}$. Since the gates are only applied in system A, the phase between ($\ket{m_1}$, $\ket{m_2^\prime}$) and between ($\ket{m_2}$, $\ket{m_1^\prime}$) will cancel. Only a CZ gate between A and B can generate the entanglement. 

On the other hand, introducing local composite measurement gates in A can reduce entanglement between A and B. As we explained in Sec.~\ref{sec:QA_meas_algorithm}, local measurements will force the bit-strings to take the same value at that site (See Fig.~\ref{fig:SWAP}). If $\ket{m_1}$ and $\ket{m_2}$ become the same in system A, it can add $4^{L}/4^{2L}=1/4^{L}$ to $\mbox{Tr}\rho_A^2$, where the $4^L$ in the numerator is  contributed by the degree of freedom in regime B. Increasing $\mbox{Tr}\rho_A^2$ by $1/4^L$ can reduce $S_A^{(2)}$ slightly. This process can be continued until all pairs of $\ket{m_1}$ and $\ket{m_2}$ become the same, i.e., the system A is completely purified. Therefore the entanglement dynamics has
\begin{align}
    S_A^{(2)}(t)=-\log_2 \frac{M(t)}{4^L},
    \label{eq:EE_purification}
\end{align}
where $M(t)$ is the number of pairs of bit-strings $A_1$ and $A_2$ which are the same at time $t$. Here $A_1$ and $A_2$ are the components of the bit-strings in subsystem A of $\ket{m_1}$ and $\ket{m_2}$ respectively. There are in total $4^L$ pairs of $A_1$ and $A_2$. At $t=0$, $M(t)=2^L$ and will increases monotonically in time under the dynamics of purification.

By following the above argument, we realize that the purification dynamics are related to a simple classical stochastic process in which the bit-strings $A_1$ and $A_2$ undergo the same random unitary and measurement dynamics. The difference between $A_1$ and $A_2$ can be characterized by the Hamming distance
\begin{align}
    D(t)=\sum_{i=1}^L|A_{1,i}-A_{2,i}|.
\end{align}
If $A_1$ and $A_2$ are initially different, then under the pure unitary evolution, $D(t)$ will be nonzero forever. For instance, consider the two-qubit CNOT gate, if the first qubit is the control gate, this can give rise to the following update rule:
\begin{align}
    \ket{00}\to \ket{00},\ket{01}\to\ket{01},\ket{10}\to\ket{11},\ket{11}\to\ket{10}.
\end{align}
In contrast, the measurement at site $i$ will force $A_1$ and $A_2$ to take the same value at site $i$ and can effectively reduce $D(t)$.
The competition between the unitary evolution and measurement leads to a non-equilibrium phase transition in this classical bit-string model. Interestingly, as we demonstrate in Appendix \ref{sec:bit_string}, this classical transition belongs to the DP universality class. When $p<p^{\text{DP}}_c$, $\overline{D(t)}/L$ approaches a finite constant (for sub-exponential time scales) and when $p\geq p^{\text{DP}}_c$, $\overline{D(t)}/L\rightarrow 0$ with a finite rate. Here $\overline{\bullet}$ denotes the ensemble average and $p^{{\text DP}}_c$ is the critical point for the classical DP transition. At the critical point, $\overline{D(t)}$ is a universal scaling function. In our purification dynamics, at $t=0$, $\overline{D}=L/2$; under time evolution, $\overline{D(t)}\sim t^{-0.159}$.

As we have shown in Eq.~\eqref{eq:EE_purification}, a pair of bit-strings $A_1$ and $A_2$ can contribute to $S^{(2)}_A$ only when their Hamming distance $D=0$. When $p>p^{\text {DP}}_c$, it typically takes $\log L$ time for $D$ between a different pair of bit-strings to become zero and therefore the quantum purification time for system A is $\log L$. On the other hand, when $p<p^{\text {DP}}_c$, it takes an exponentially long time for $D$ to decay to zero. Therefore, in the QA circuit model, there exists a quantum purification phase transition which occurs at $p=p_c^{\text{DP}}$ and corresponds to a change in scaling of the purification time.

 We numerically verify that the above ideas, derived from the classical bit-string model, directly apply to the full quantum model by simulating the entanglement dynamics of the QA Clifford circuit depicted in Fig.~\ref{fig:Clifford_purification_3}. The unitary dynamics is very simple and is composed of two types of CNOT gates. In principle, we could also introduce phase gates like the CZ gate in the unitary dynamics. However, as we explained before, the phases for $\ket{m_1},\ket{m_2^\prime}$ and $\ket{m_2},\ket{m_1^\prime}$ will be cancelled. They have no influence on the entanglement dynamics. Notably, since this is a Clifford circuit, we can perform large-scale simulation by directly employing the Gottesman–Knill algorithm \cite{Aaronson_2004,gottesman1998heisenberg}. We find that there is a phase transition at $p_c=0.137$. This is equal to $p_c^{\text{DP}}$ for the classical DP transition of the bit-string model (See Appendix \ref{sec:bit_string} for the details of the classical model). When $p>p_c$, $\overline{S_A^{(2)}}$ decays exponentially in time, indicating that it purifies at a constant rate independent of system size. When $p=p_c$, it decays algebraically with exponent $\approx 0.63$. We use the MC algorithm described in Sec.~\ref{sec:QA_meas_algorithm} to compute $\overline{S_A^{(2)}}$ and we find consistent results (See Fig.~\ref{fig:puri_comparision}). In addition, we define $P(t)$ as the probability for two bit-strings to be the same at time $t$. We compute $-\log(P(t))$ as a function of $t$ and find that it also decays as a power law in time.

In Ref.~\onlinecite{gullans2019purification}, they studied the purification dynamics in a random hybrid Clifford circuit which does not belong to the QA circuit class. At the critical point, they found that $\overline{S_A^{(2)}}=F(t/L)$ which decays as $L/t$ at early times. This indicates that the random Clifford circuit has dynamical exponent $z=1$. Such scaling behavior was further investigated in Ref.~\onlinecite{li2020conformal}, where they showed that this model is equipped with two dimensional conformal symmetry. However, it is known that the DP universality class has an anisotropy between the spatial and time directions with $z=1.581$. Therefore there is no conformal symmetry in the hybrid QA circuit. This is indeed consistent with our finding that $z=1/0.63$. This result is further supported in the data collapse of different system sizes in Fig.~\ref{fig:puri_collapse}, in which we show that $\overline{S_A^{(2)}}=F(t/L^z)$. We believe that such scaling behavior is universal for all the hybrid QA circuits.

Notably, among all the entanglement measures, the second R\'enyi entropy appears to be rather uniquely accessible in experiments \cite{Brydges_2019,Greiner_nature}.  Due to the simple structure of the hybrid QA circuit model in Fig.~\ref{fig:puri_cartoon}, in which the unitary evolution is composed of only CNOT gates, it has the potential to be realized in experiments in an efficient way. 
It would be interesting to experimentally observe this purification transition and compare the critical exponents with that of the DP universality class.

\section{Entanglement transition}
In this section, we explore the entanglement transition in the hybrid QA circuit model. We start from the product state $\ket{\psi_0}$ with $L$ qubits and investigate the entanglement dynamics of a continuous subsystem A. 

Under generic random unitary QA dynamics, we approach a random phase state, i.e.,
\begin{align}
    \ket{\psi}=\frac{1}{\sqrt{2^L}}\sum_m e^{i\theta_m}\ket{m}
    \label{eq:rp_state}
\end{align}
where $\theta_m$ is a random phase uniformly distributed in $[0,2\pi]$. This state has near maximal entanglement entropy and can be understood from Eq.~\eqref{eq:EE}. As shown in Fig.~\ref{fig:SWAP}, if  $\ket{m_1}$ and $\ket{m_2}$ are the same in subsystem A, we have $\Theta_{1,m}=-\Theta_{2,m}$. There are in total $2^{L_A}\times 4^{L_B}$ such pairs which can contribute $1/2^{L_A}$ to $\mbox{Tr}\rho_A^2$. Each of the remaining states will contribute a random phase and when we sum them up, their mean contribution is zero. Therefore for the random phase state we have $S_A\approx L_A$ when $L_A<L/2$.

We can also analyze the entanglement dynamics under the unitary evolution in terms of the classical bit-string dynamics. For $\ket{\psi_0}$, applying a unitary gate fully within region A will not increase the entanglement. This is because under this unitary gate, the phases from  $\ket{m_1}=A_1B_1$ and $\ket{m_2^\prime}=A_1B_2$ will cancel with each other. This is also true for $\ket{m_2}=A_2B_2$ and $\ket{m_1^\prime}=A_2B_1$. For the rest of the discussion, we will focus on the pair $(\ket{m_1},\ket{m_2^\prime})$ ( $(\ket{m_2},\ket{m_1^\prime})$ can be analyzed in the same way). Only a gate which crosses the boundary between A and B can increase the entanglement. In addition, if $B_1$ is different from $B_2$, this boundary gate can induce a difference in A between $\ket{m_1}$ and $\ket{m_2}^{\prime}$. We can characterize this difference by introducing the function $h(x,t)$, where
\begin{align}
    h(x,t)=|m_1(x,t)-m_2^\prime(x,t)|.
\end{align}

\begin{figure}[t]
 \includegraphics[width=.9\columnwidth]{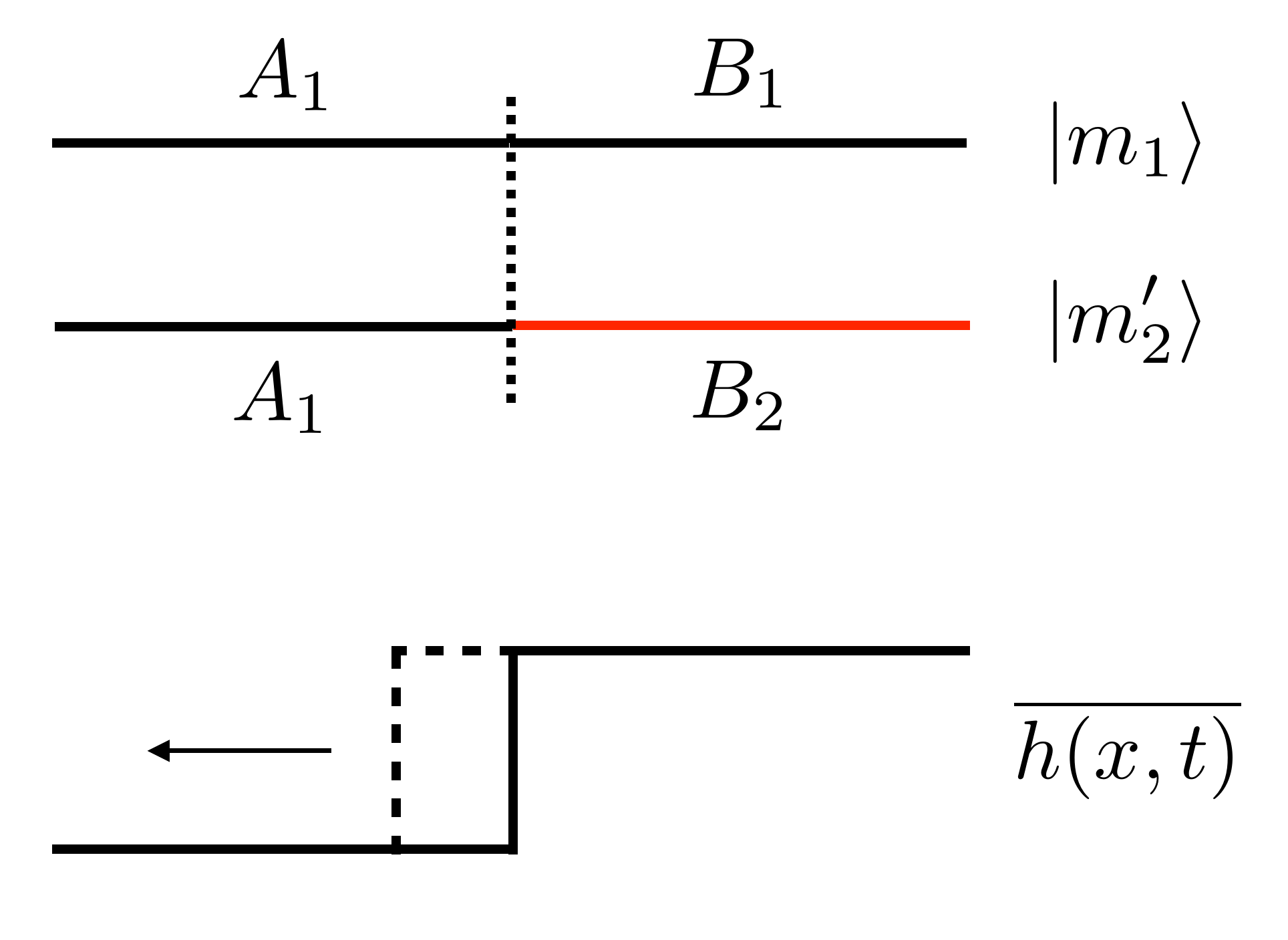}
\caption{ The dynamics of $\overline{h(x,t)}$ under unitary time evolution. $h(x,t)$ is defined as the difference between two random bit strings $\ket{m_1}$ and $\ket{m_2^\prime}$. Under unitary time evolution, the front of $\overline{h(x,t)}$ moves to the left with a constant velocity. In this schematics, we ignore the possible broadening of the wavefront. }  
\label{fig:front_dyn}
\end{figure}

At the initial time, in the classical bit-string model the two strings are identical in A.  Applying unitary gates across the boundary of $A$ can flip the boundary bit (we take it to be the ``rightmost" bit) in $A$:  hence the bit-string difference $h(x,t)$ is schematically $00\cdots 01$.  Now, under the unitary dynamics, we can have $01 \rightarrow 01,10,11$. Therefore, there will be a ``wavefront" moving to the left, behind which there will be a finite density of 1s.
As illustrated in Fig.~\ref{fig:front_dyn}, with the help of the unitary gates in regime A, the front of $\overline{h(x,t)}$ will move to the left with a constant velocity. The regime behind the front will be ``scrambled".  We provide detailed evidence of this behavior for $\overline{h(x,t)}$ in Appendix \ref{app:DP}. 
 In the full quantum evaluation of $S^{(2)}$, bit-strings $\ket{m_1}$ and $\ket{m_2^\prime}$ which are different in this regime can carry with them a random relative phase. In the most general case, this relative phase averages to zero. Therefore, only pairs of bit-strings which are exactly the same in this regime contribute to the entanglement. This therefore implies that we have $S_A\sim  l(t)$ where $l(t)$ is the length of this scrambled regime which grows linearly in time. 

We can generalize the above idea to the hybrid dynamics with composite measurements. When $p>p_c^{
\text{ DP}}$, with frequent measurement,  we force the bit-string to take the  same value at every site on which a measurement is performed.  Since the measurement rate is sufficiently fast, even if there are unitary dynamics, $\ket{m_1}$ and $\ket{m_2^\prime}$ become the same after a finite amount of time. The front of $h(x,t)$ can only spread to a finite regime in A, with $l(t)$ being finite, when we take into account the unitary gates across the boundary of A that can appear in the middle of the circuit. Therefore the steady state entanglement entropy is bounded by the maximal value of $l(t)$ and satisfies the area law. In contrast, when $p<p_c^{\text{DP}}$, with infrequent measurement, $h(x,t)$ can spread over the entire regime in A and we expect the entanglement entropy to satisfy volume law scaling. 

Based on the above argument, we make the following conjecture. At the critical point, the entanglement dynamics has the form
\begin{eqnarray}
\overline{S^{(2)}(t)} = \alpha_1 \log(t).
\end{eqnarray}
The steady state has the entanglement entropy form
\begin{eqnarray}
\overline{S^{(2)}(L_A)} = \alpha_2 \log(L_A).
\label{eq:steady_EE}
\end{eqnarray}
This scaling form has also been observed in the ground state of one dimensional spin chain models with $z>1$\cite{Bravyi_2012,Chen_Fradkin_2017,Fisher_1994,Refael_2004}. The coefficients $\alpha_1$ and $\alpha_2$ are non-universal but the ratio between them must obey \begin{equation}
\frac{\alpha_2}{\alpha_1}=z.
\end{equation} Furthermore, around the critical point ($p<p_c$), the steady state $S^{(2)}(L_A)$ satisfies
\begin{align}
    \overline{S^{(2)}(L_A,p)}&=\alpha_2\log\xi_{\perp}+\frac{L}{\xi_{\perp}}\nonumber\\
    &=\alpha_2\log\xi_{\perp}+L(p-p_c)^{\nu_{\perp}},
\end{align}
where $v_{\perp}=1.0969$.

To verify the above conjecture, below we study a Clifford circuit and a non-Clifford circuit and check the entanglement scaling in them numerically. The Clifford QA circuits allow for an exact simulation of extremely large systems so we can accurately determine the value of all critical exponents. We then verify that the same universal behavior holds in the more general non-Clifford QA circuits.


\subsection{Clifford circuit}
\begin{figure}[t]
 \includegraphics[width=.7\columnwidth]{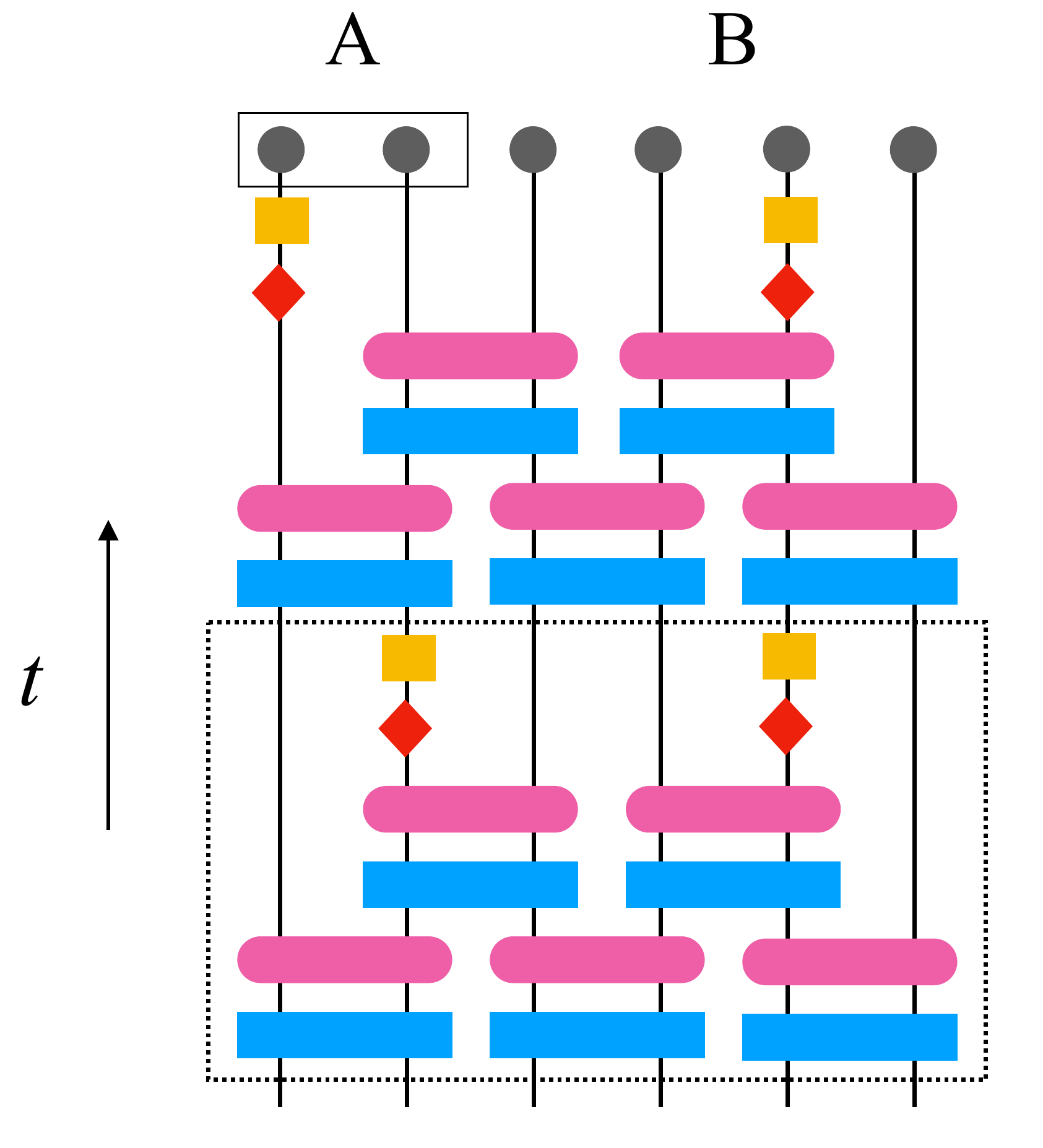}
\caption{ The cartoon for a QA Clifford circuit constructed from CNOT gates, CZ gates, Z measurement gates and H gates. The unitary dynamics has a brick wall structure. There are two types of CNOT gates and we apply them randomly with equal probability. The Z measurement gate (accompanied with the H gate) is applied randomly at each time step with probability $p$. The dashed box denotes one time step. We start with a product state polarized in the x direction and study the entanglement dynamics and the scaling of the steady state entanglement. } 
\label{fig:Clifford_EE}
\end{figure}

We first consider a Clifford circuit model described in Fig.~\ref{fig:Clifford_EE} and compute the entanglement entropy using the Gottesman–Knill algorithm \cite{Aaronson_2004,gottesman1998heisenberg}. The unitary evolution is composed of CNOT gates and CZ gates. Different from the purification dynamics in Fig.~\ref{fig:Clifford_purification_3}, we now need to include the phase gate in the unitary evolution to generate entanglement. 

\begin{figure}[hbt]
\centering
\subfigure[]{\label{fig:EE_p} \includegraphics[width=.9\columnwidth]{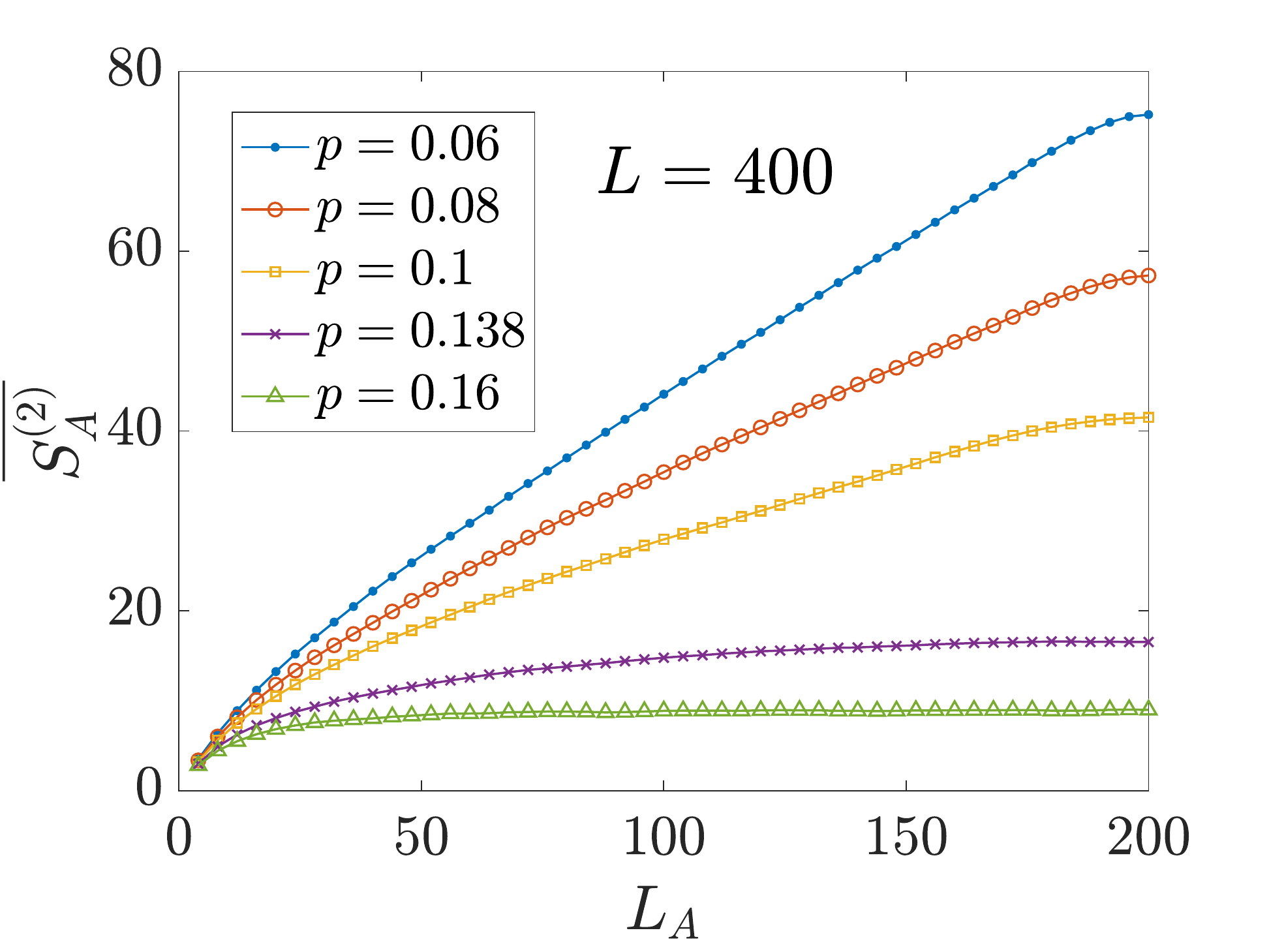}}
\subfigure[]{\label{fig:EE_scaling} \includegraphics[width=.9\columnwidth]{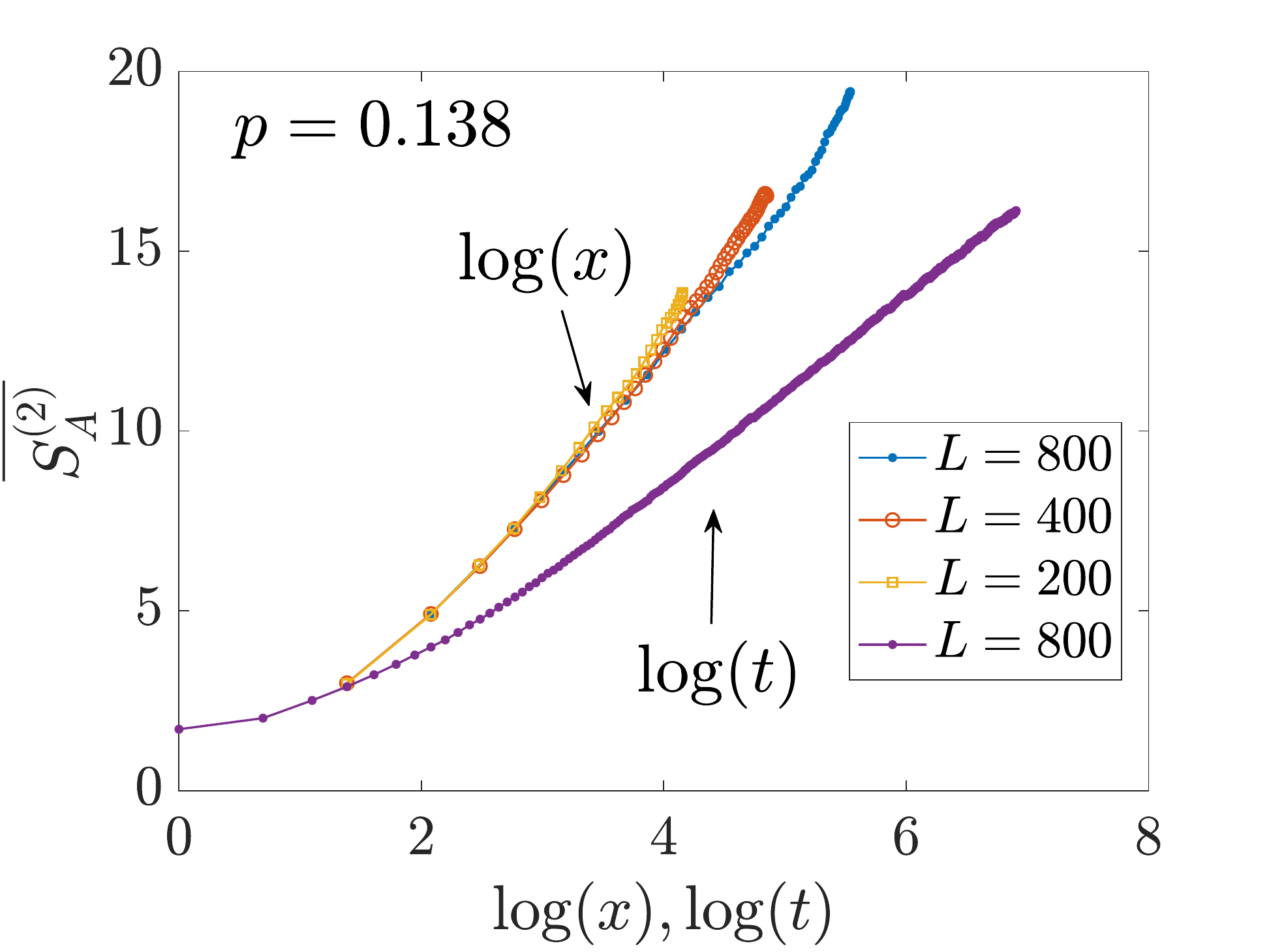}}
\caption{ (a) The steady state $\overline{S_A^{(2)}}$ vs $L_A$ at different $p$. (b) The first three curves are steady state $\overline{S_A^{(2)}}$ vs $\log x$ for various system sizes, where $x\equiv\sin(\pi L_A/L)L/\pi$. These three curves collapse into a single curve when $\log x$ is small. Due to finite size effects, these curves are bending up as we increase $\log x$, i.e., they are not exactly at the critical point when $L$ is small. Such behavior will disappear when we increase the system size. The slope of the $L=800$ curve is $4.033$. The last purple curve is the entanglement dynamics for half of the system. The slope of the curve is $2.655$. The ratio between these two slopes is $1.519$. In the calculation, we take periodic boundary conditions.}
\label{fig:EE_result}
\end{figure}

We present the steady state result in Fig.~\ref{fig:EE_p} and we show that there exists a volume law phase. As we increase $p$, the volume law coefficient decreases and disappears around $p=0.138$. When $p>0.138$, it enters into the area law phase with $S_A^{(2)}$ independent of the subsystem size. At the critical point, the data collapse on the 1d chain with periodic boundary conditions in Fig.~\ref{fig:EE_scaling} suggests that
\begin{align}
    \overline{S_A^{(2)}}=\alpha_2 \log\left[\frac{L\sin(\frac{\pi L_A}{L})}{\pi}\right],
\end{align}
consistent with the result in Eq.\eqref{eq:steady_EE}.
We further study the entanglement dynamics and 
we find $
    \overline{S_A^{(2)}}=\alpha_1\log t$.
Both scaling behaviors have also been observed in other non-unitary quantum dynamics at a critical point. However, different from those critical points which have emergent two dimensional conformal symmetry with $\alpha_1=\alpha_2$ \cite{li2020conformal,Chen_2020}, in our non-unitary QA circuit, $\alpha_1\neq \alpha_2$. Indeed the ratio is $\alpha_2/\alpha_1=1.519\approx z$.

\begin{figure}[hbt]
\centering
\subfigure[]{\label{fig:MI_eta} \includegraphics[width=.9\columnwidth]{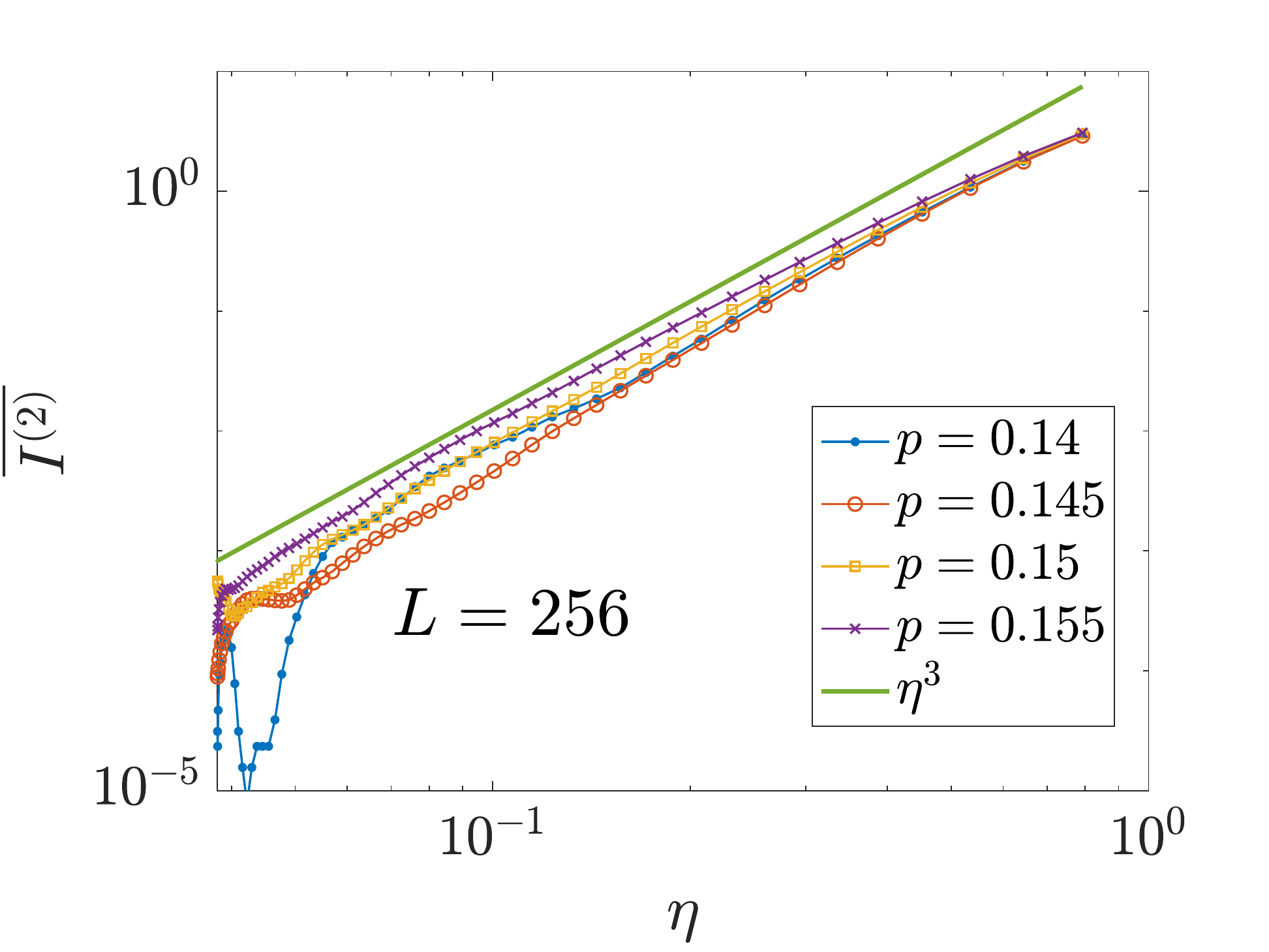}}
\subfigure[]{\label{fig:MI_p} \includegraphics[width=.9\columnwidth]{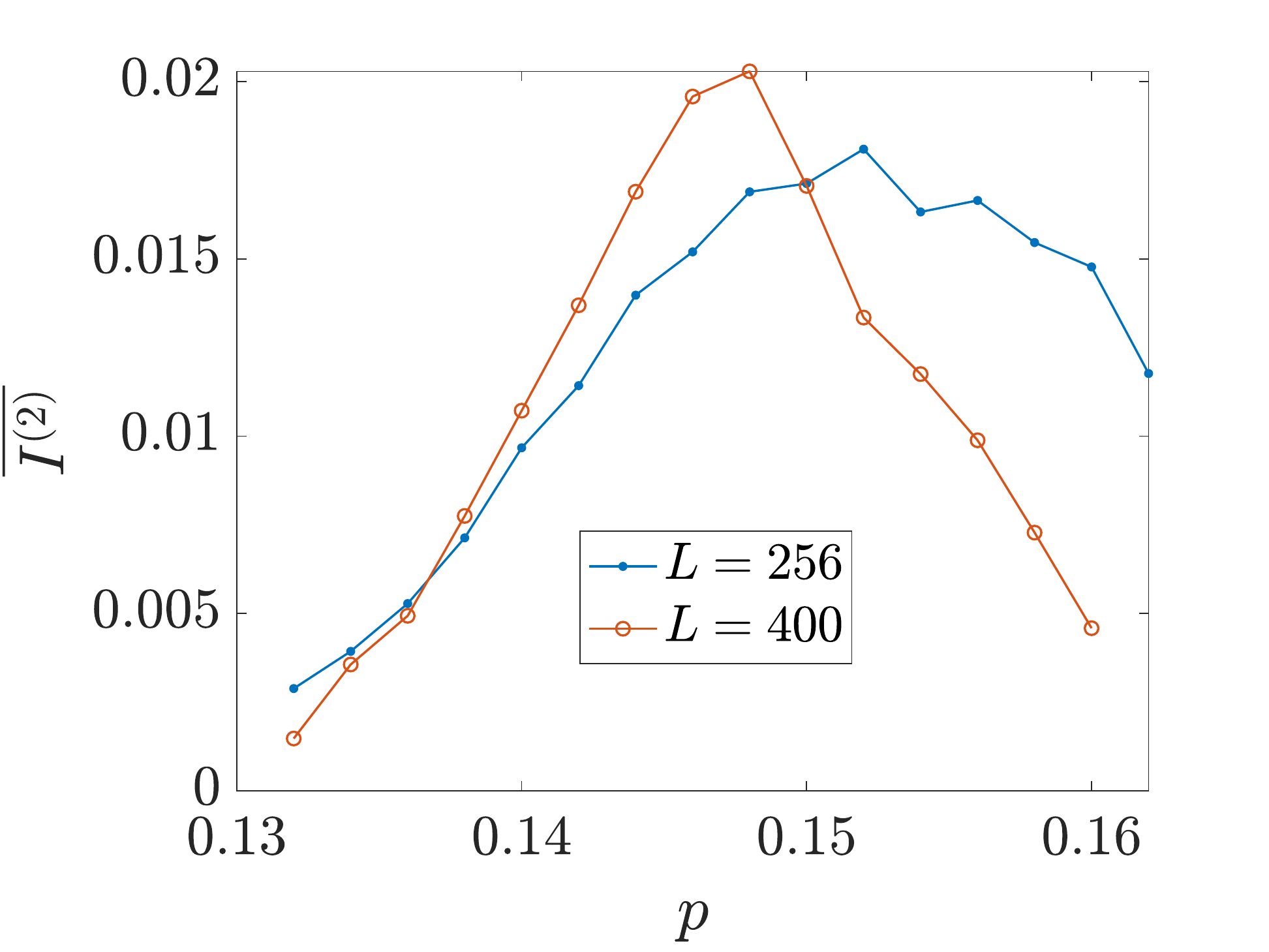}}
\caption{ (a) The mutual information $\overline{I^{(2)}}$ as a function of cross ratio at various $p$ close to $p_c^{{\text DP}}$ for the Clifford QA circuit. (b) $\overline{I^{(2)}}$ between two antipodal regions A and B with length $L_A=L_B=L/8$. In the calculation, we take periodic boundary conditions.}
\label{fig:EE_result}
\end{figure}

We further compute the mutual information  around the critical point. The mutual information between two intervals is defined as,
\begin{align}
    I^{(2)}=S_A^{(2)}+S_B^{(2)}-S_{A\cup B}^{(2)}.
\end{align}
We plot it as a function of cross ratio and present it in Fig.~\ref{fig:MI_eta}. Here the cross ratio is defined as 
\begin{align}
    \eta=\frac{x_{12}x_{34}}{x_{13}x_{24}},\ \text{with}\ x_{ij}=\sin\left(\frac{\pi}{L}|x_i-x_j|\right).
\end{align}
Since the mutual information is very small, numerically it is difficult to compute it accurately. In the simulation, we consider a smaller system with $L=256$ and take two intervals with $L_A=L_B=16$. We move these intervals on the circle and we find that around this critical point
\begin{align}
   \overline{I^{(2)}}\sim \eta^\Delta,\ \text{with}\ \Delta=3, 
\end{align} 
when $\eta\ll 1$. This result implies that $\overline{I^{(2)}}\sim 1/r^6$ for two small distant intervals. It is interesting to see that this exponent is larger than $\Delta=2$ which was observed in both the hybrid random Clifford circuit model and hybrid random Haar circuit model \cite{Li_2019,Nahum_Skinner_2020}. It further indicates that our critical wave function is less correlated than what was obtained in these hybrid circuit models with conformal symmetry. Notice that the critical point can be identified as the peak of the mutual information in Fig.~\ref{fig:MI_p} \cite{Li_2019} and is slightly larger than $p_c^{\text{DP}}$. This is due to the finite size effect \footnote{Compared with the purification transition which has $p_c=0.137$, it seems that there is a stronger finite size effect in the entanglement transition.}. As we increase  $L$, the peak of the mutual information moves to the left slightly. We expect that this peak will coincide with $p_c^{\text{DP}}$ in the thermodynamic limit.

\begin{figure}[t]
 \includegraphics[width=.9\columnwidth]{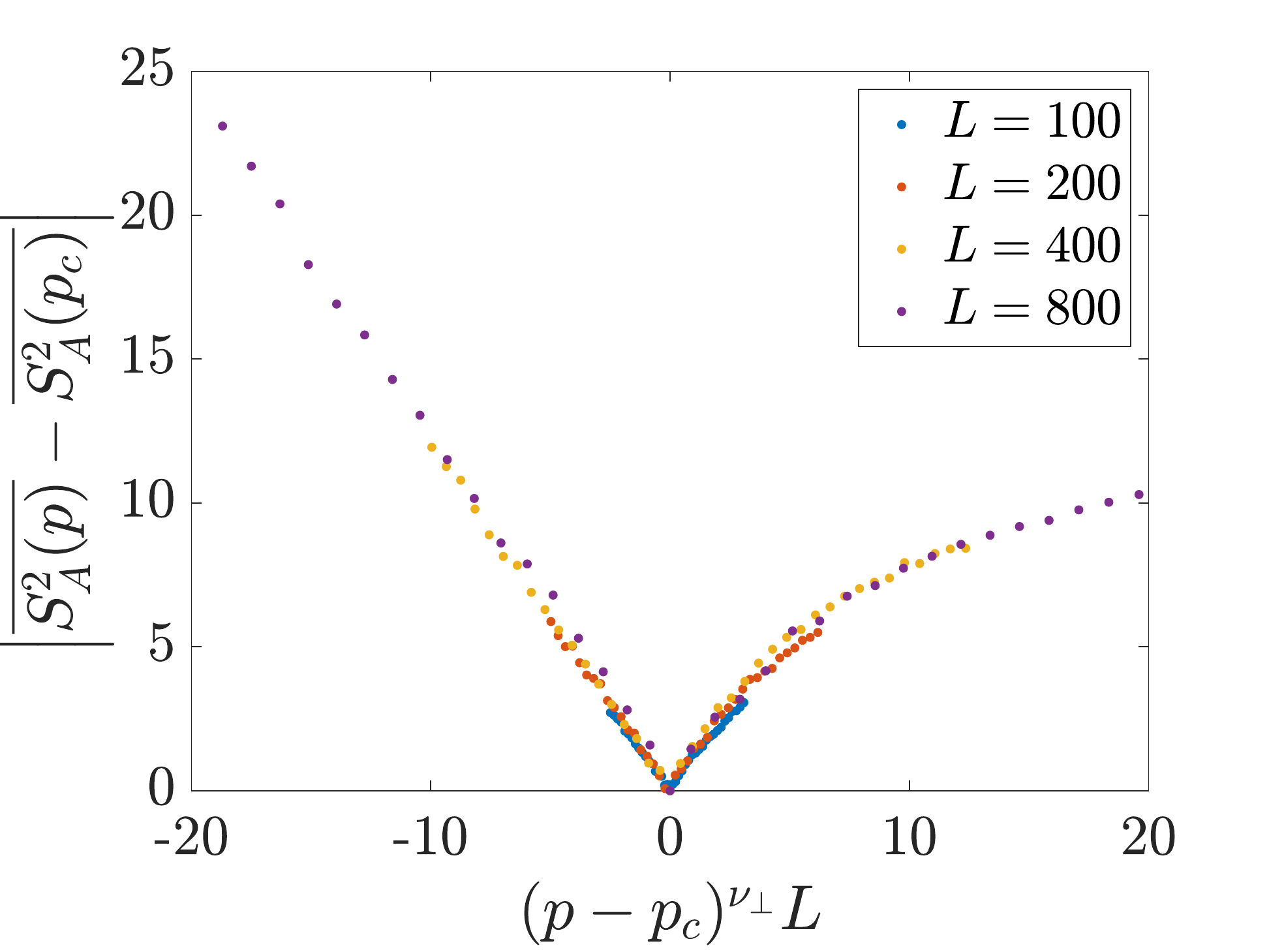}
\caption{ Data collapse of $\overline{S_A^{(2)}}$ around the critical point for the  Clifford QA circuit. Here we take $L_A=L/4$. All these data points of different $L$ collapse to a single scaling function $\overline{S_A^{(2)}(p)}-\overline{S_A^{(2)}(p_c)}=F(L/\xi_{\perp})$. When $p<p_c$, $F(L/\xi_{\perp})\sim L(p-p_c)^{\nu_{\perp}}$ with $\nu_{\perp}=1.0969$. We use the value of $v_\perp$ from the DP universality class and find a very good data collapse. } 
\label{fig:EE_collapse}
\end{figure}

\subsection{Non-Clifford circuit}

\begin{figure}[t]
 \includegraphics[width=.8\columnwidth]{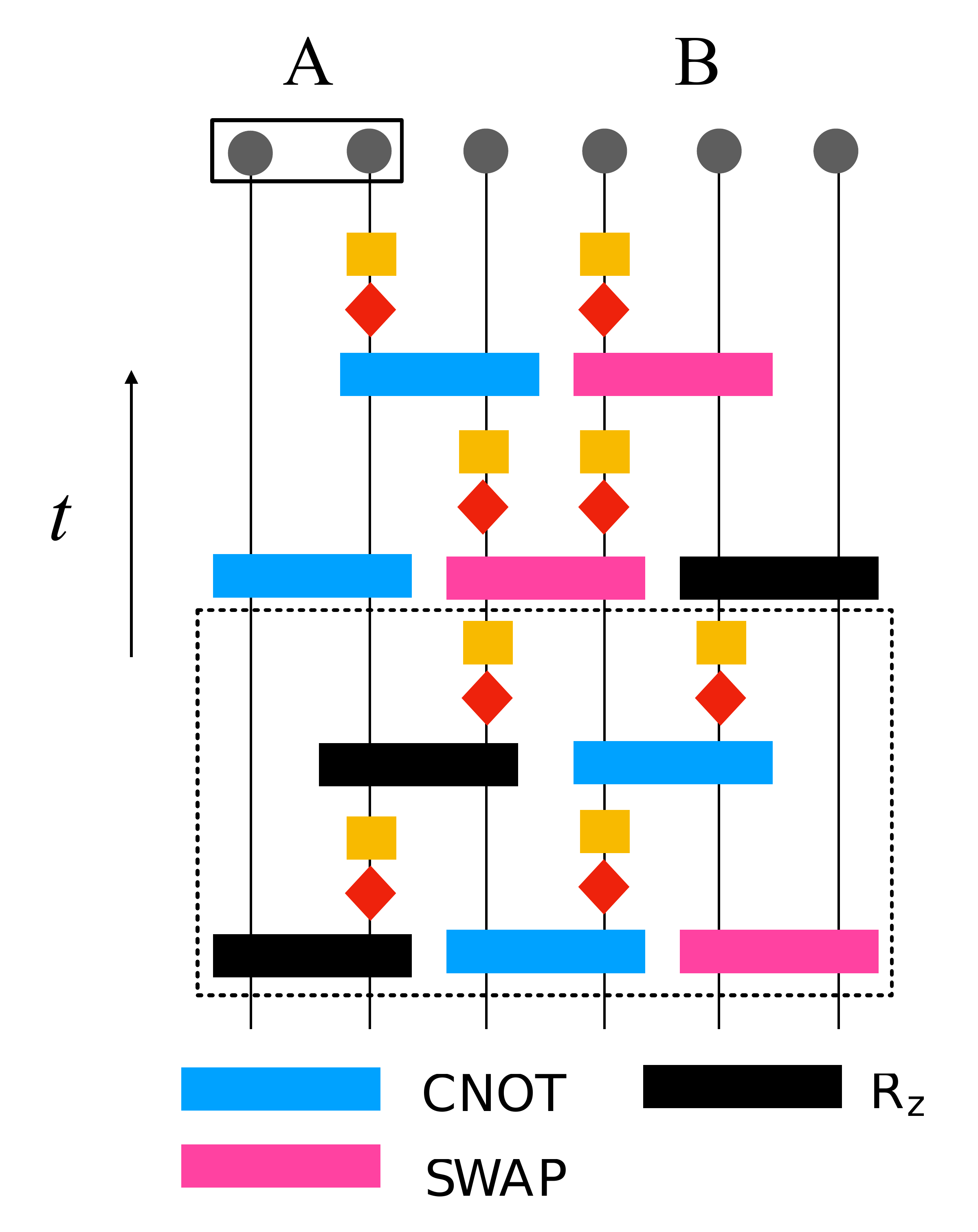}
\caption{ The cartoon for a hybrid QA circuit. The unitary evolution is constructed from CNOT, SWAP, and R$_z$ gates. Different from the Clifford circuits we studied in this paper, in each time step, we apply measurement gates after a single layer of unitary operators. The corresponding bit-string model has a smaller value $p_c^{\text{DP}}=0.053$.} 
\label{fig:Non_Clifford}
\end{figure}

We now consider hybrid QA circuits which include gates outside of the Clifford set.   Specifically, we consider a circuit composed of the randomly chosen 2-site gates CNOT, SWAP, R$_Z=e^{i(\theta_1 Z_1 +\theta_2 Z_2 +\theta_3 Z_1Z_2)}$, with random phases $\theta_i\in[0,2\pi]$ which are uniformly distributed (See Fig.~\ref{fig:Non_Clifford}). We again apply these gates to the fully polarized initial states $\ket{\psi_0}$. In the absence of measurement, the steady state will approach a random phase state as in Eq.~\eqref{eq:rp_state} which exhibits maximal volume law entanglement scaling. Interestingly, the structure of entanglement in this more general QA state is distinct from that of Clifford wave functions \cite{iaconis2020quantum}. For example, at late times in generic QA circuits without measurement, fluctuations of the bipartite entanglement entropy of the wave function are qualitatively more strongly suppressed compared to Clifford wave functions. In addition, unlike with Clifford wave functions, the full entanglement spectrum in this generic case exhibits Gaussian unitary ensemble (GUE) random matrix statistics. Nevertheless, we find that the universal properties at the phase transition in this more general class of circuits are again related to the classical DP critical point.

\begin{figure}[h]
\includegraphics[scale=0.45]{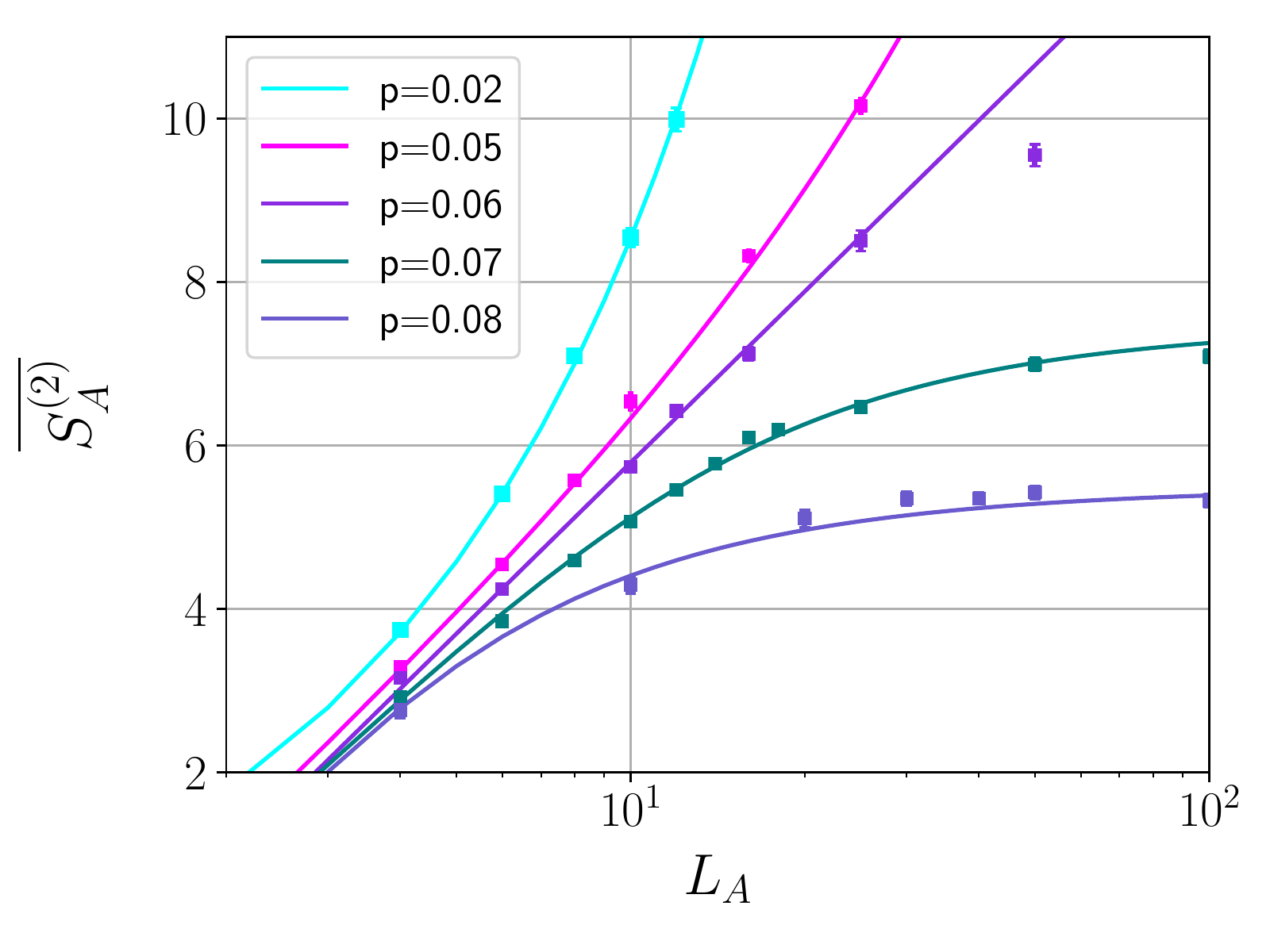}
\caption{$\overline{S_A^{(2)}}$ vs $L_A$ on the semi-log scale for the generic (non-Clifford) QA circuit. The system size is $L=200$. The critical point appears to be $p_c\approx 0.06$. For $p<p_c$, the $\overline{S_A^{(2)}}$ curves possess a clear volume law component. For $p>p_c$, the curves appear to saturate to a constant area law at large $L_A$. }
\label{fig:QA_SL}
\end{figure}

Since we are studying non-Clifford dynamics, we must employ the MC algorithm of Sec.~\ref{sec:QAreview}.  This limits the amount of entanglement we can resolve in the simulation. Still, we are able to study the entanglement growth in much larger circuits than for the Haar random case.  
In Fig.~\ref{fig:QA_SL}, we plot the late time entanglement as a function of subsystem size $S^{(2)}(L_A)$, for circuits with $L=200$ qubits. We see that there appears to be a phase transition around $p_c\approx 0.06$. This value is close to $p_c^{\text{DP}}=0.053$ of the corresponding classical bit-string model. The small difference is due to the finite system size. For $p>p_c$, the entanglement saturates to a constant ``area law'' entanglement for large subsystem size $L_A$.  On the other hand, for $p<p_c$ the entanglement appears to grow up to the largest entropy values we are able to resolve. In this regime, the best fit to the data shows a linear ``volume law'' component to the entanglement growth.  Close to the transition, we find that the entanglement grows like $
\overline{S_A^{(2)}(L_A)} = \alpha_2 \log(L_A) $
with $\alpha_2 \approx 3.0$.  Note that slightly above the transition, the entanglement correlation length is large and we must measure relatively large regions $L_A\geq 50$ before we see the saturation to an area law entanglement.  

\begin{figure}[h]
\includegraphics[scale=0.45]{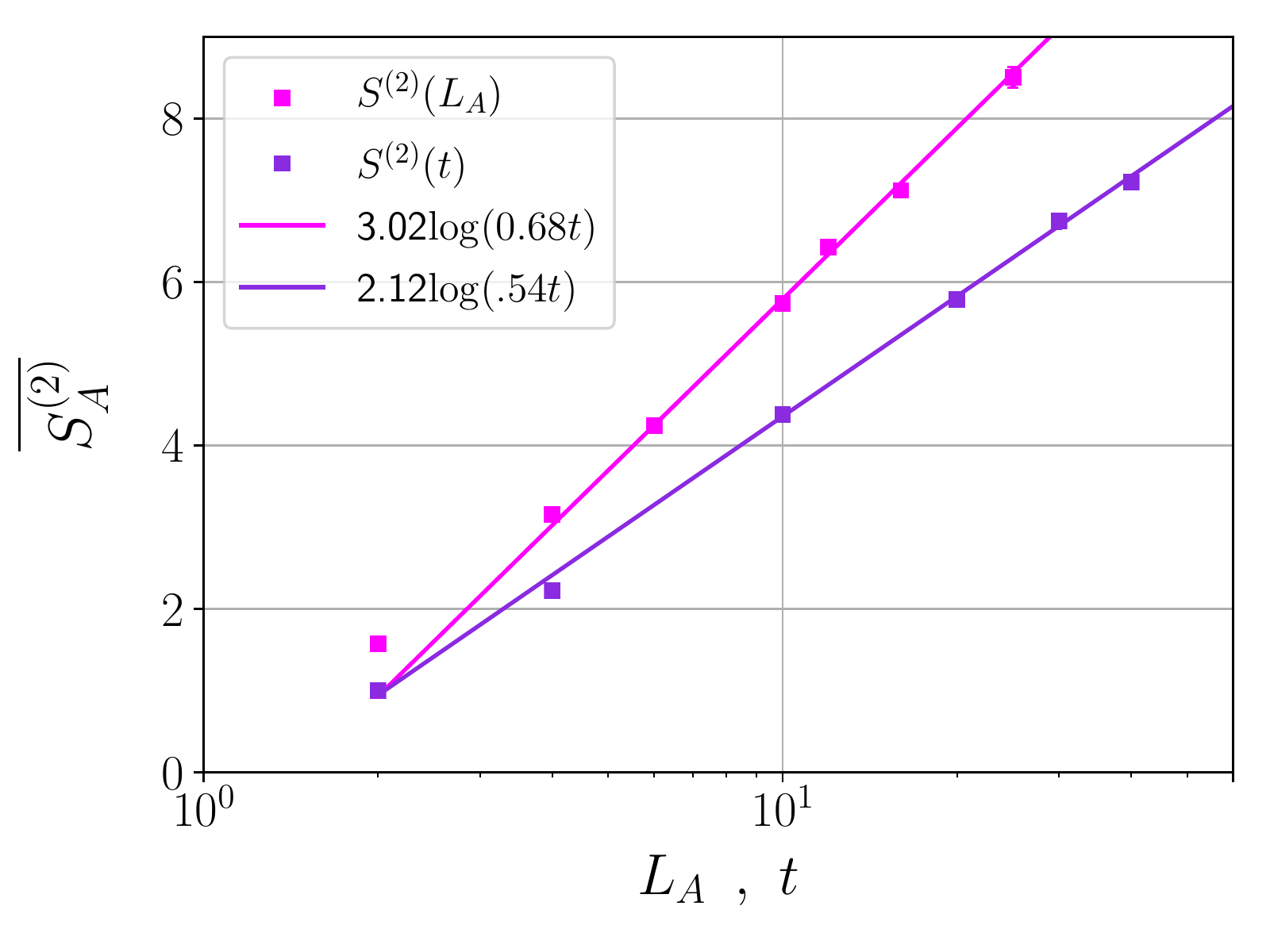}
\caption{ The entanglement scaling at the critical point $p_c=0.06$ for the non-Clifford QA circuit. For both curves, the system size is $L=200$, with periodic boundary conditions. The pink curve is $\overline{S^{(2)}}(L_A)$ vs $L_A$. The purple curve is $\overline{S^{(2)}}(t)$ vs $t$ with $L_A=100$. }
\label{fig:QA_St}
\end{figure}

In Fig.~\ref{fig:QA_St}, we show the growth of entanglement as a function of circuit time. We again measure the entanglement for circuits with $L=200$ sites, for subsystem size $L_A=100$. Near the transition, we again see that entanglement grows like
$\overline{S^{(2)}(t)} = \alpha_1 \log(t)$.
We find that $\alpha_1 \approx 2.1$.  Notice that the ratio $\alpha_2/\alpha_1 \approx 1.43$, is close to the dynamical critical exponent for the (1+1)d DP universality class. 

We also compute the mutual information between two intervals in this non-Clifford QA model.
Since the mutual information is very small, this calculation is considerably harder to perform using the MC algorithms than the simple entanglement calculation. Consequently, we are only computing the mutual information in circuits with up to $L=42$ sites and with periodic boundary conditions. Note that this still outperforms other algorithms for studying quantum dynamics. 

\begin{figure}[t]
\centering
   \subfigure[]{\label{fig:mi_f} \includegraphics[width=.9\columnwidth]{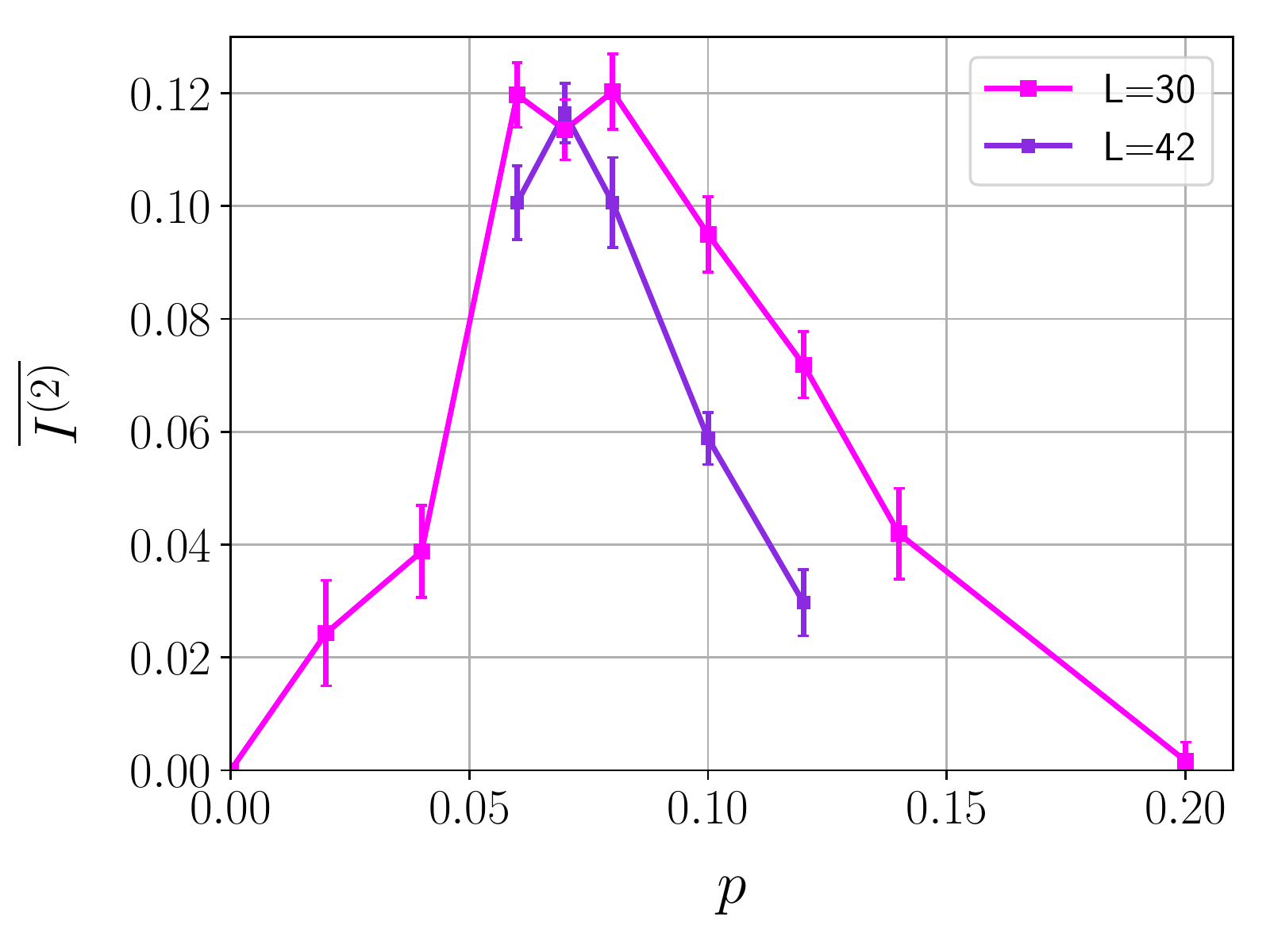}}
\subfigure[]{\label{fig:mi_scaling} \includegraphics[width=.9\columnwidth]{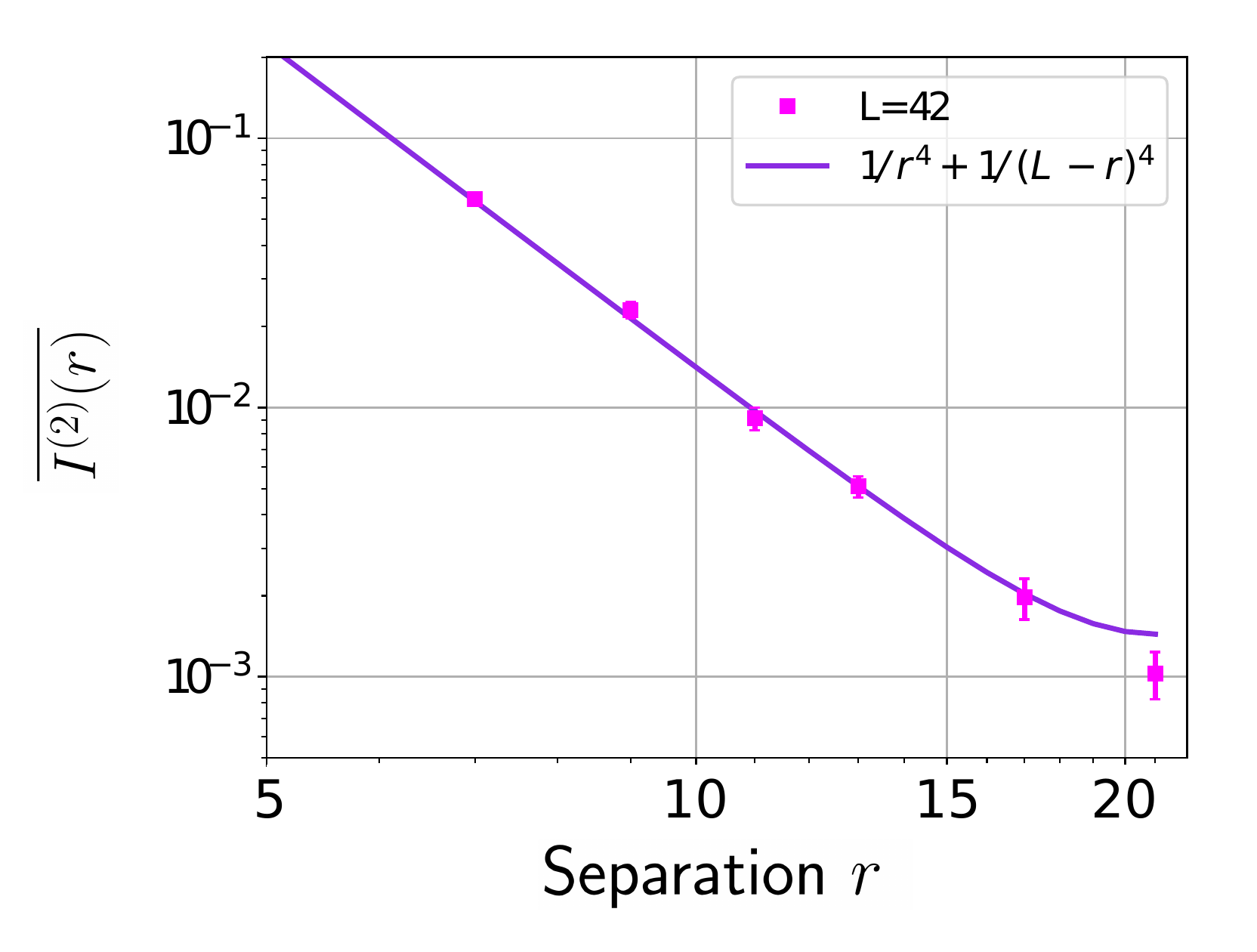}}
\caption{(a) The mutual information $\overline{I^{(2)}}$ vs $p$ for the QA circuit defined in Fig.~\ref{fig:Non_Clifford} for two antipodal regions with $L_A=L_B=L/6$.  It shows a clear peak near $p=0.07$. (b) At $p=0.07$, we plot $\overline{I^{(2)}}$ vs separation distance between regions A and B, with $|A|=|B|=3$, with $L=42$ and periodic boundary conditions. }
\label{fig:QA_MI}
\end{figure}

In Fig.~\ref{fig:mi_f}, we plot $\overline{I^{(2)}}$ as a function of measurement probability $p$, for circuits with $L=30$ and $L=42$ sites. We choose regions $A$ and $B$ to be centered around antipodal points on the circuit and let  $|A|=|B|=L/6$.  We see there is a distinct peak in the mutual information near $p=0.07$.  This measurement rate is slightly higher than the transition we see in the $L=200$ circuit, which we expect is a finite size effect.  

At the finite size transition for $L=42$ circuits, $p_c=0.07$, we also compute the mutual information as a function of separation between regions $A$ and $B$. Specifically, we set $|A|=|B|=3$ and plot $\overline{I^{(2)}(r)}$, where $r$ is the separation between the middle sites of the two regions.  We show this data in Fig.~\ref{fig:mi_scaling}, where we see a clear power law decay of the mutual information $1/r^4$.  The critical exponent in these finite size systems appears distinct from the Clifford QA case, suggesting that the critical exponent $\overline{I^{(2)}}$ is model dependent.

\section{Discussion and Conclusion}
In this paper, we study a measurement-induced entanglement phase transition in a $(1+1)$d hybrid QA circuit composed of both unitary gates and projective measurements followed by rotations. We show that the second R\'enyi entropy $S_A^{(2)}$ can be understood in terms of a classical bit-string model whose phase transition belongs to the DP universality class with $z=1.581$. We performed large scale simulations for various QA circuits and numerically observed an entanglement transition and a DP transition which occur at the same $p_c$, in agreement with our expectations. Furthermore, at the critical point, in the purification dynamics, we confirmed that the entanglement entropy of a region of size $L$ scales as a function of $L/t^{1/z}$. We also compute the entanglement dynamics from a product state and we find that $S_A^{(2)}$ grows as $\alpha_1\log t$. In addition, the steady state entanglement grows as $S_A^{(2)}=\alpha_2\log L_A$. The ratio between these two coefficients is $\alpha_2/\alpha_1=z$.

In the classical non-equilibrium dynamics, the DP universality class has been shown to be very robust. In the bit-string model, we have considered various unitary dynamics and we find that they all show the same critical behavior. This is also true for the corresponding  quantum entanglement dynamics. This motivates us to ask the following question: is there another universality class for the QA entanglement transition which is distinct from the DP universality class? The answer is yes. One way to realize this is to introduce additional symmetry. For instance, we can impose a parity-conserving constraint in the bit-string dynamics so that the number of $1/0$ bits is conserved modulo 2. Such a model has $\mathbb{Z}_2$ symmetry and belongs to a universality class distinct from the DP universality class\footnote{The critical point with $\mathbb{Z}_2$ symmetry has $z=1.76$ \cite{hinirchsen2008non}.}. This parity-conserving constraint can also be manifested in the hybrid QA model with both unitary and measurement gates. It would also be interesting to explore the phase transition in non-unitary QA circuit models with more complicated discrete or even continuous symmetries \footnote{Similar ideas have been explored in non-QA hybrid circuits \cite{sang2020measurement,lavasani2020measurementinduced,Lang_2020,Nahum_Skinner_2020}.}.

Another way to go beyond the (1+1)d DP universality class is to introduce long-range interactions in the unitary dynamics \cite{gullans2019purification,vijay2020measurementdriven,nahum2020measurement}. We have also considered such models with all-to-all interactions, and we find a measurement-driven phase transition with non-local interactions, but it is clearly not in the same universality class.  Also note that in this case, there is a purification phase transition where the phases  under the entanglement dynamics from $\ket{\psi_0}$ on either side of the transition are volume law entangled.  It would also be interesting to study the more complicated case of power law interactions.  In addition, we can also introduce non-local measurement that involves multiple qubits.  We leave the study of possible phase transitions in these models for future study.

 It is widely believed that there is a generic entanglement phase transition in the hybrid quantum circuit, or monitored Hamiltonian dynamics. Therefore it is not surprising to observe a phase transition in the hybrid QA circuit. Indeed, our model retains much of the phenomenology of the previously studied CFT critical points, such as a logarithmic scaling of the entanglement entropy and a power law decay of the mutual information.  What is most informative about our model is the bit-string picture and the mapping to the classical DP dynamics, as a simple cartoon for understanding this phase transition. Further, using this picture, we can more rigorously prove that a phase transition in the second R\'enyi entropy exists. Such an analytic understanding does not exist for the non-QA Haar random and Clifford hybrid circuits.  However, in the hybrid QA circuit, if we introduce non-QA unitary gates, the classical bit-string picture immediately breaks down. In models with non-QA gates and local interaction, we observe that $z=1$ is restored (See Appendix \ref{sec:non-QA} for more discussion on a Clifford circuit). In fact, these systems have two dimensional conformal symmetry, just as the random Clifford circuit studied in Ref.~\onlinecite{li2020conformal}. 
 Among all known examples of measurement based entanglement transitions, the hybrid QA circuit is the only known model without conformal symmetry. We are not sure whether the QA circuits are peculiar in this sense or if there exist far more undiscovered universality classes beyond the QA circuits with measurement.
 In the future, it would also be interesting to explore the possible crossover between these different universality classes.

\section*{Acknowledgements}
We thank Michael Gullans, Timothy Hsieh and Yaodong Li for helpful discussions. XC thanks Matthew Fisher, Yaodong Li and Andreas Ludwig for collaborations on related projects. AL was supported by a Research Fellowship from the Alfred P. Sloan Foundation.  This material is based in part (J.I.) upon work supported by the Air Force Office of Scientific Research under award number FA9550-20-1-0222.  This work is partially supported (J.I.) by the National Science Foundation under Grant Number 1734006. J.I. is supported by a Simons Investigator Award to Leo Radzihovsky from the Simons Foundation.

\begin{appendix}
\section{Directed percolation}
\label{app:DP}\label{sec:bit_string}
Classical stochastic processes can exhibit interesting non-equilibrium phase transitions.  Many different stochastic models contain phase transitions which belong to the directed percolation (DP) universality class. Near such a transition point, certain properties of these models exhibit universal behavior and possess identical critical exponents. Unlike with the theory of (undirected) percolation, the DP universality class has not been exactly solved. Many of the critical exponents are only obtained through large scale numerical simulation, yet have been shown to be extremely robust. In this appendix, we consider two simple examples in (1+1)d: the bond directed percolation and the bit-string model related to the QA circuit. More details of the DP universality class can be found in Ref.~\onlinecite{hinirchsen2008non}.

\subsection{Bond directed percolation}
\begin{figure}[hbt]
\centering
   \subfigure[]{\label{fig:BDP_seed} \includegraphics[width=.9\columnwidth]{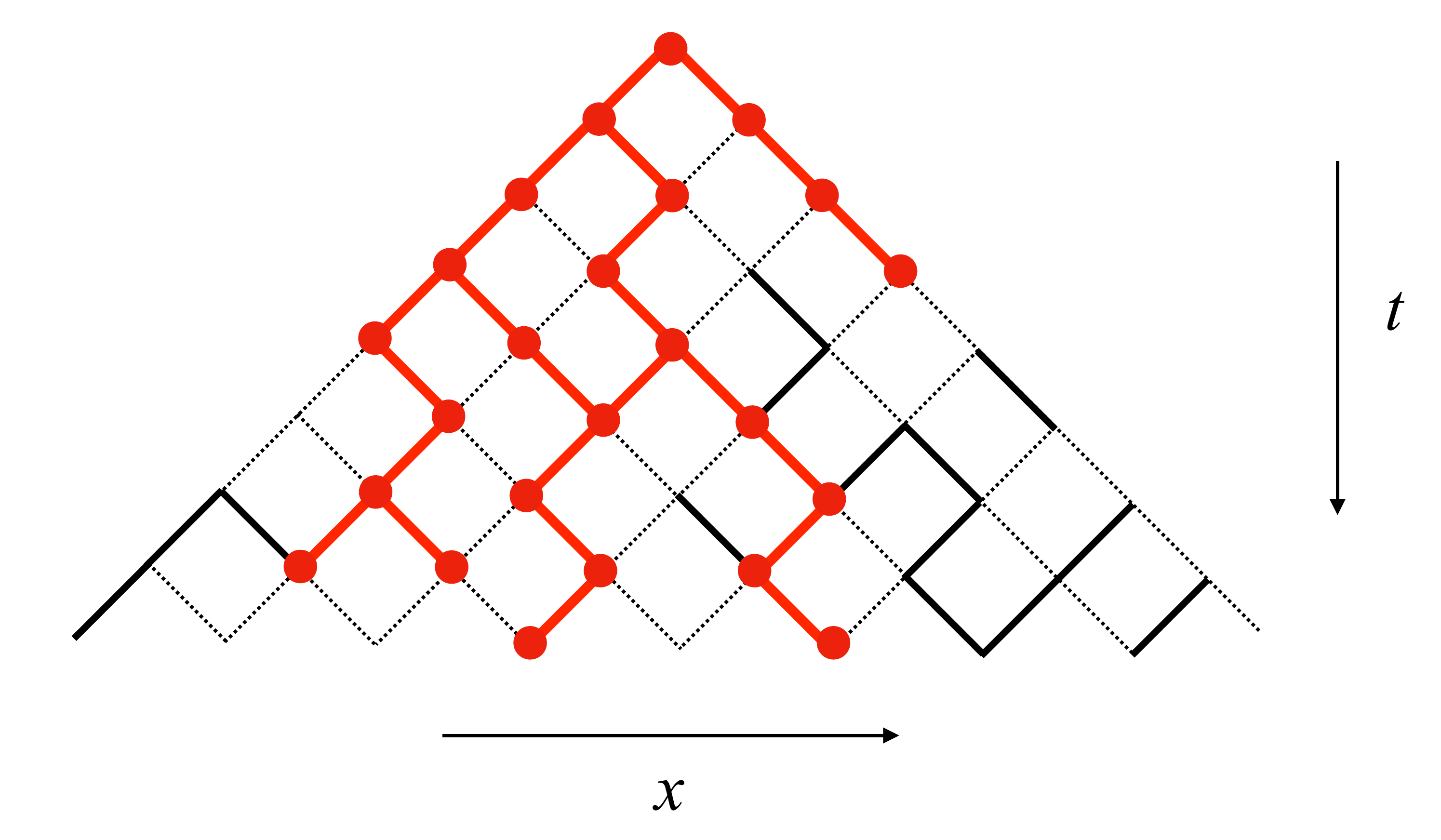}}
\subfigure[]{\label{fig:BDP_rule} \includegraphics[width=.9\columnwidth]{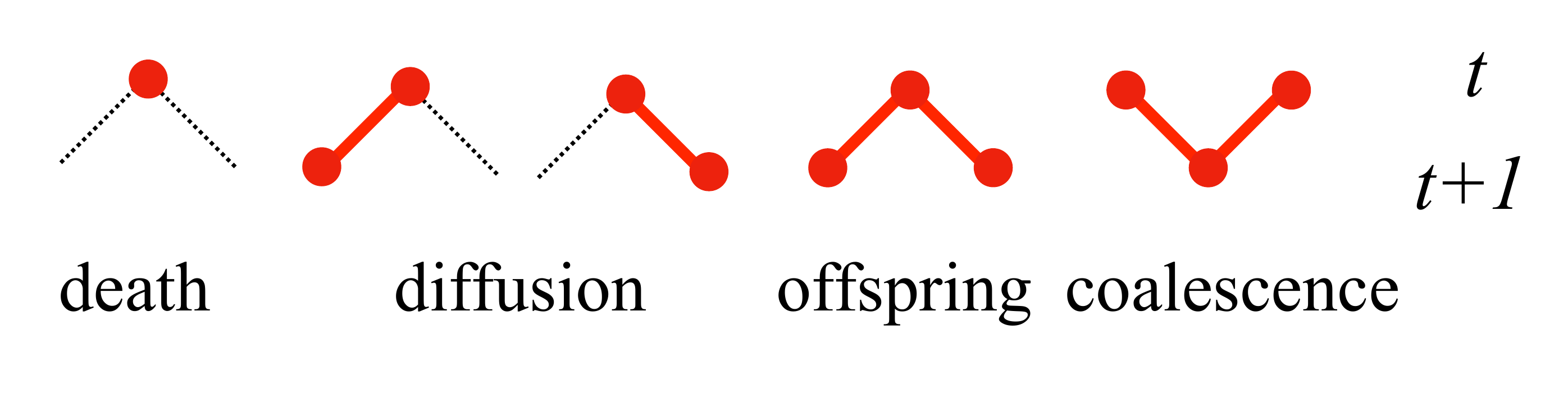}}
\subfigure[]{\label{fig:BDP_uniform} \includegraphics[width=.9\columnwidth]{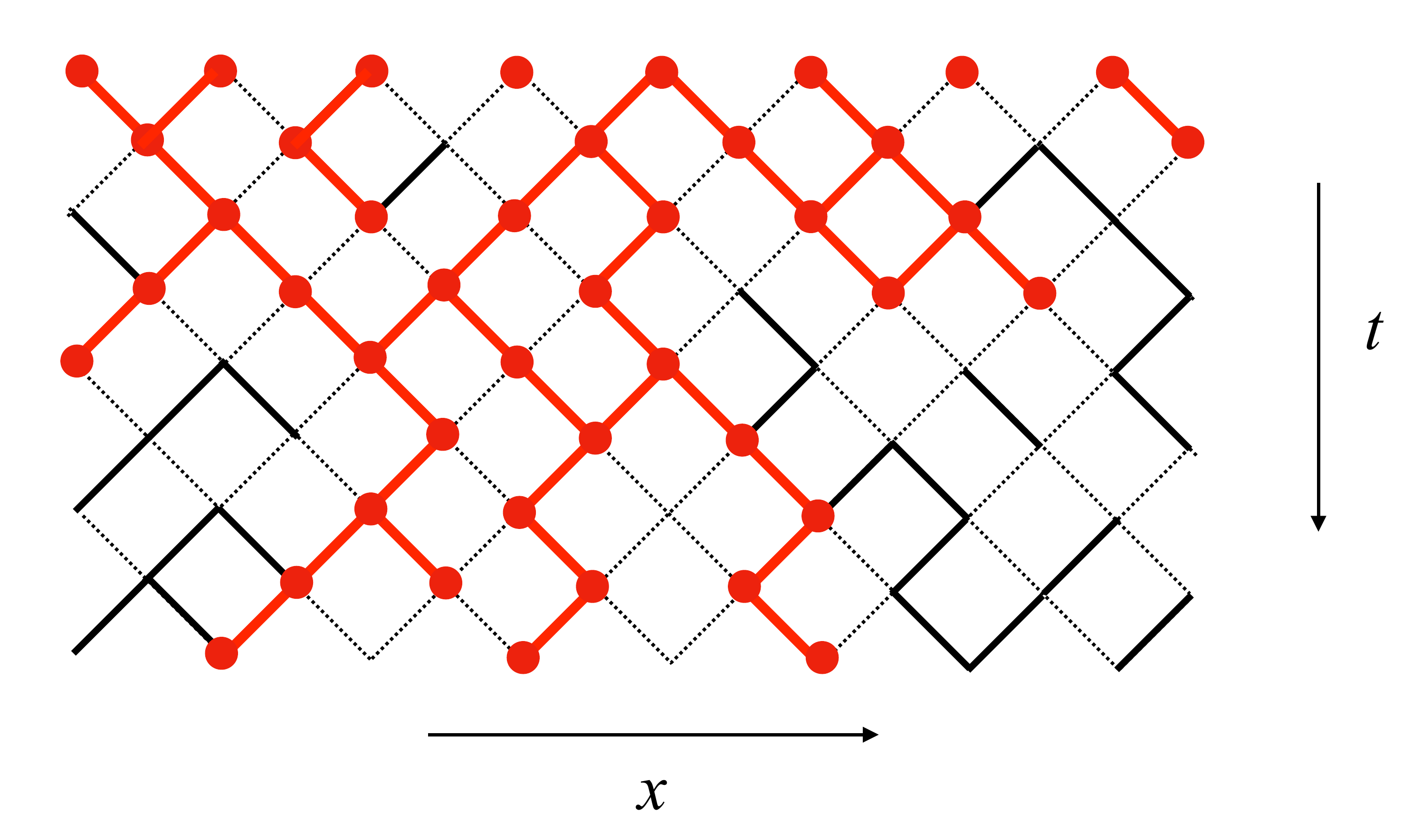}}
\caption{The bond DP model on the tilted square lattice, where the dashed line and solid line denote the broken bond and connected bond respectively. (a) The initial condition with a single site; (b) The bond DP dynamics can be interpreted as the  diffusion-reaction process involving the update rule described in Fig.~\ref{fig:BDP_rule}. (c) The initial condition with fully occupied sites.} 
\label{fig:BDP}
\end{figure}

The bond DP model is one of the simplest realizations of a model which is known to belong to the DP universality class. As illustrated in Fig.~\ref{fig:BDP_seed}, the bond DP model consists of a space-time lattice in which certain bonds are randomly blocked with probability $p$. Here, the vertical direction is the time direction and the horizontal direction is the spatial direction. Occupied sites on this lattice then evolve according to an update rule which is described in Fig.~\ref{fig:BDP_rule}. If we start with an initial condition with only a single occupied seed (See Fig.~\ref{fig:BDP_seed}), then as time evolves this seed can spread out along the connected bonds. When $p<p_c$, this spreading has a linear light cone structure in space-time and the particle number at each time step is $N(t)\sim t$. In contrast, when $p>p_c$, the penetration depth is finite and the cluster formed by all the connected bonds remains finite. Therefore, in the long time limit, $N(t\to\infty)=0$. 

In this bond DP model, the final steady state particle number density $\rho\equiv N(t\to\infty)/L$ can be treated as an order parameter which characterizes the phase transition. $\rho$ changes continuously as we vary $p$: it takes a finite value when $p<p_c$ and is zero when $p\geq p_c$. Therefore a non-equilibrium phase transition takes place at the critical threshold $p_c$. 

Exactly at $p_c$, starting from the single seed initial state, at early times, the averaged particle number is $\overline{N(t)}\sim t ^{\Theta}$ with $\Theta\approx 0.302$ \cite{hinirchsen2008non}. We could also consider other initial states, such as the fully occupied state with $\rho=1$ (See Fig.~\ref{fig:BDP_uniform}). For this initial state,  $\overline{N(t)}$ decays as a power law function, i.e., $\overline{N(t)}\sim t^{-\beta/\nu_{\parallel}}=t^{-0.1595}$ \cite{hinirchsen2008non}. These two exponents $\Theta$ and $\beta/\nu_{\parallel}$ are universal quantities for the DP universality class. 

At the critical point, the structure of the directed percolation cluster is anisotropic. For the correlation length, the corresponding critical exponent in the vertical (temporal) direction has $\nu_{\parallel}=1.7338$ while the critical exponent in the horizontal (spatial) direction has $\nu_{\perp}=1.0969$. Therefore the dynamical exponent $z=\nu_{\parallel}/\nu_{\perp}$ is not equal to 1. This is different from the (undirected) percolation critical point, which has $\nu_{\parallel}=\nu_{\perp}=4/3$ and two dimensional conformal symmetry at $p_c$.

\subsection{Bit-string model}

\begin{figure*}[t]
\centering
\subfigure[]{\label{fig:BS_purification} \includegraphics[width=.9\columnwidth]{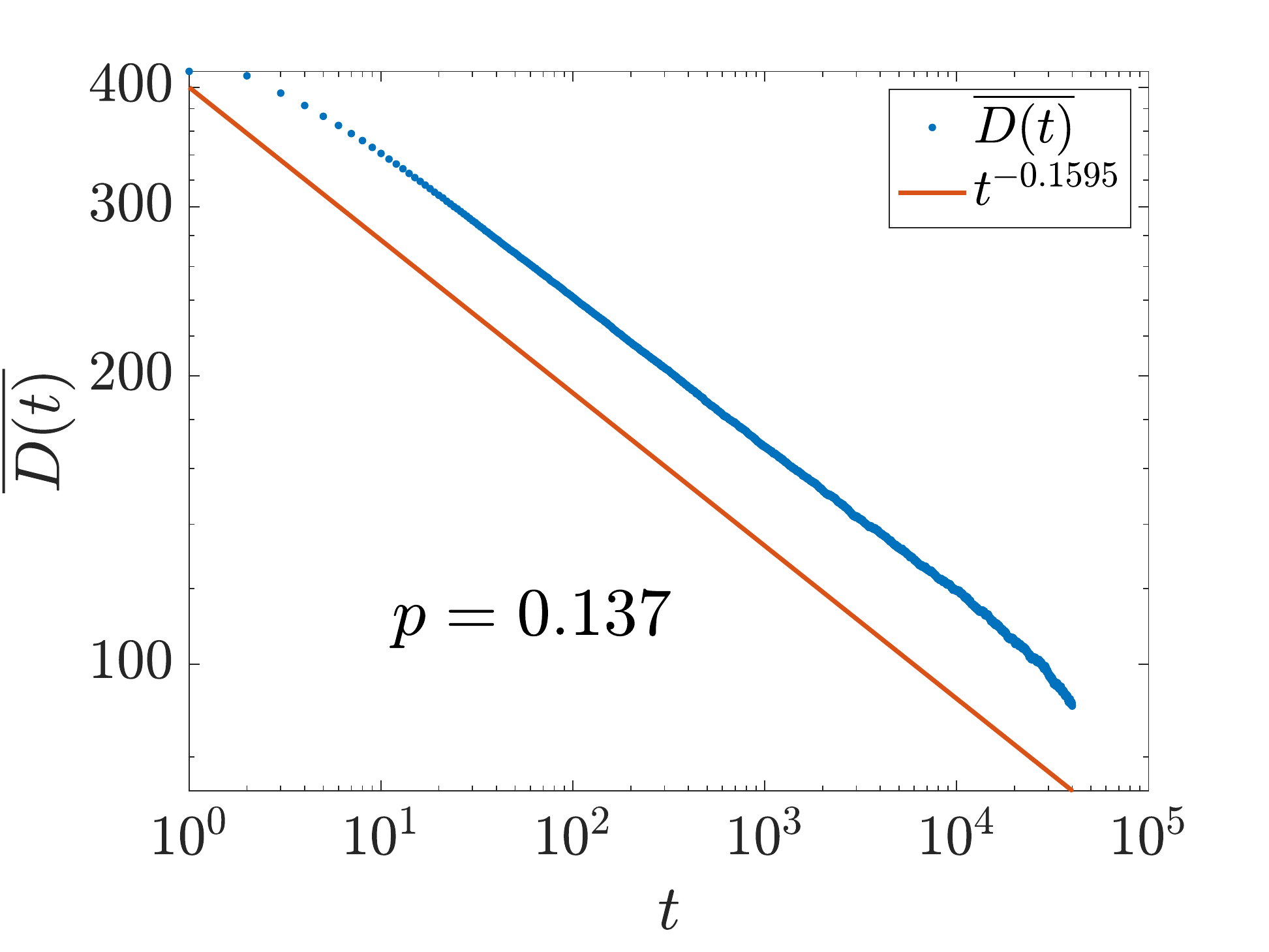}}
\subfigure[]{\label{fig:BS_spreading} \includegraphics[width=.9\columnwidth]{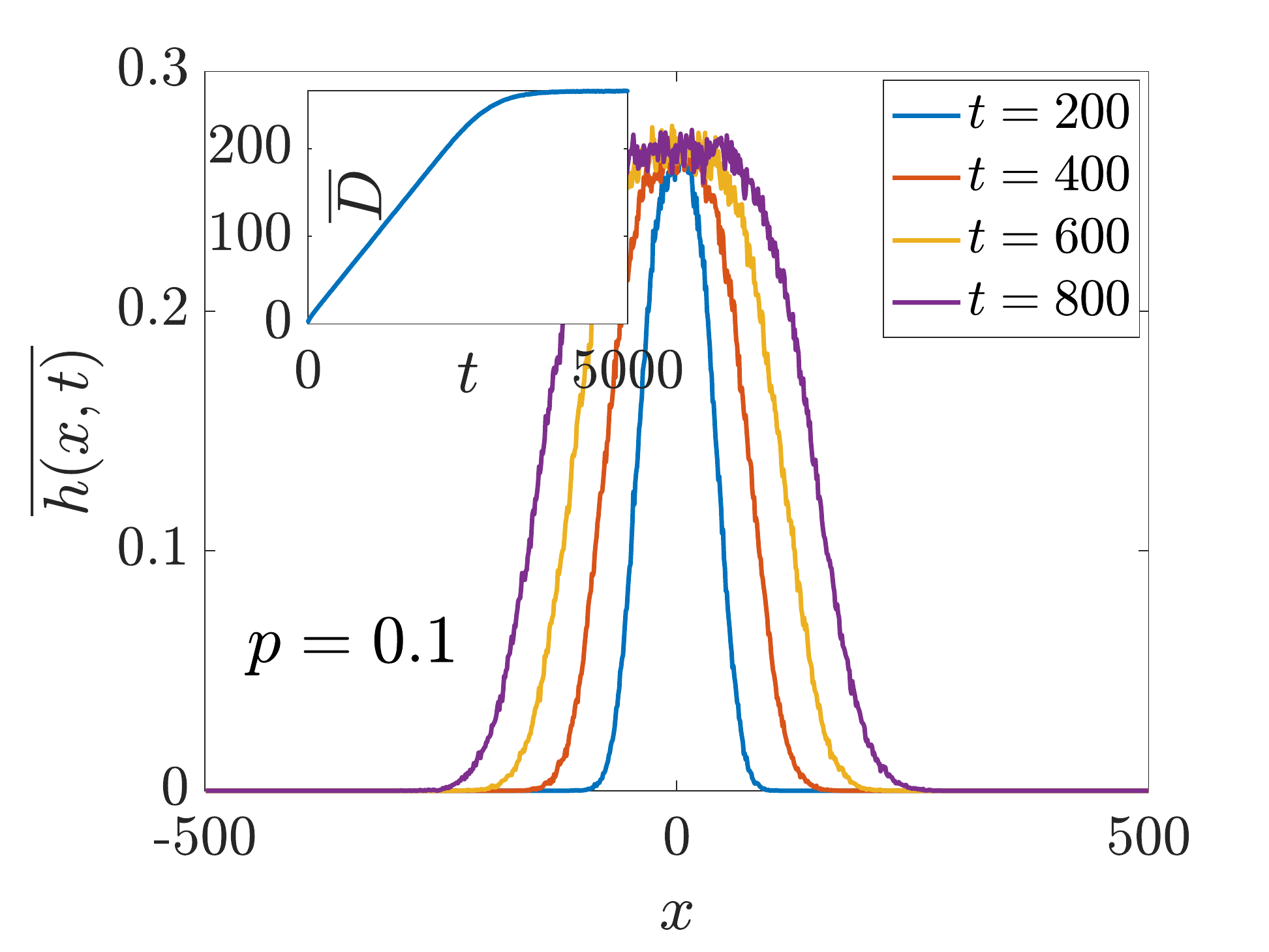}}
\subfigure[]{\label{fig:BS_spreading_collapse} \includegraphics[width=.9\columnwidth]{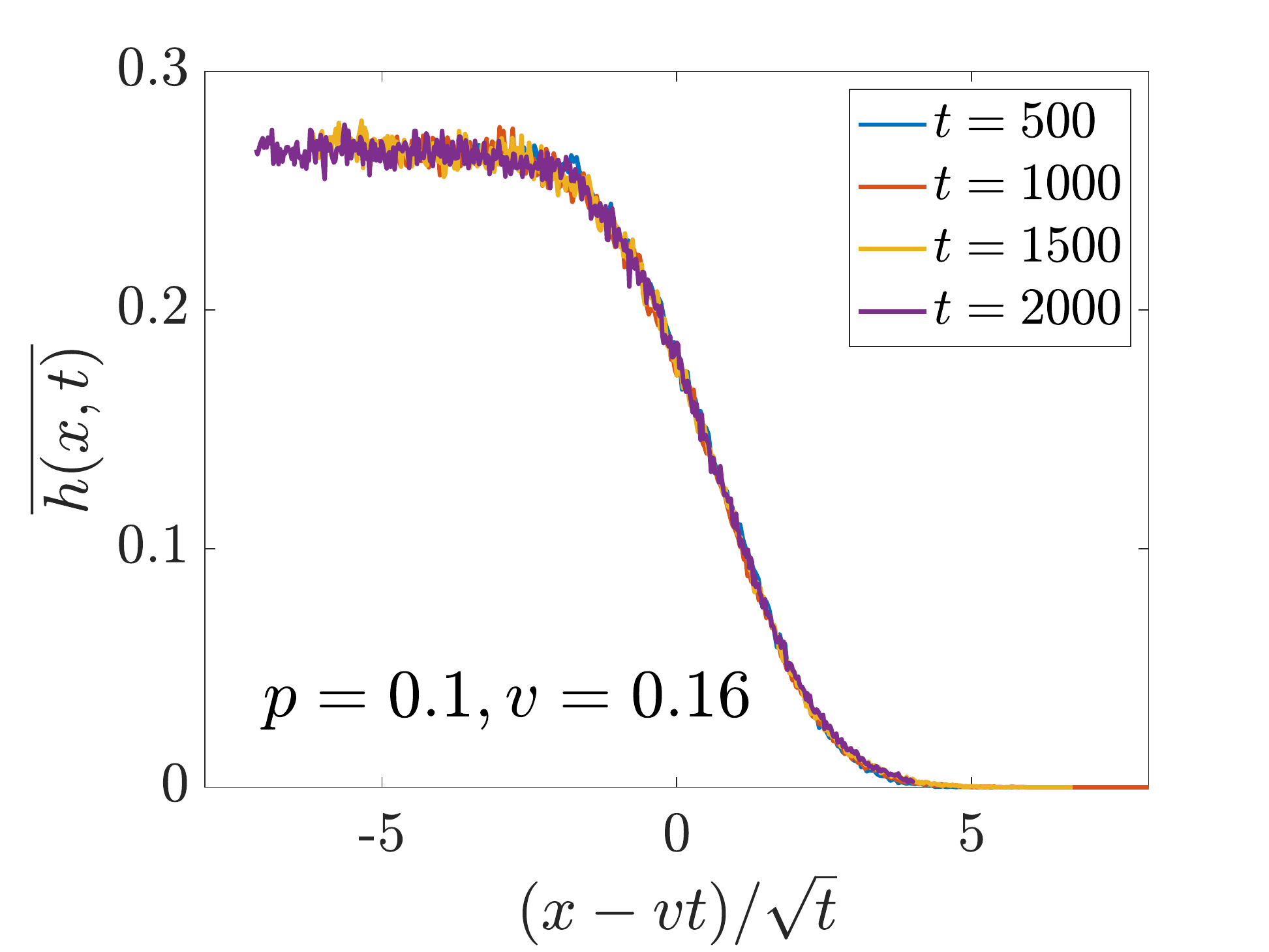}}
\subfigure[]{\label{fig:BS_seed} \includegraphics[width=.9\columnwidth]{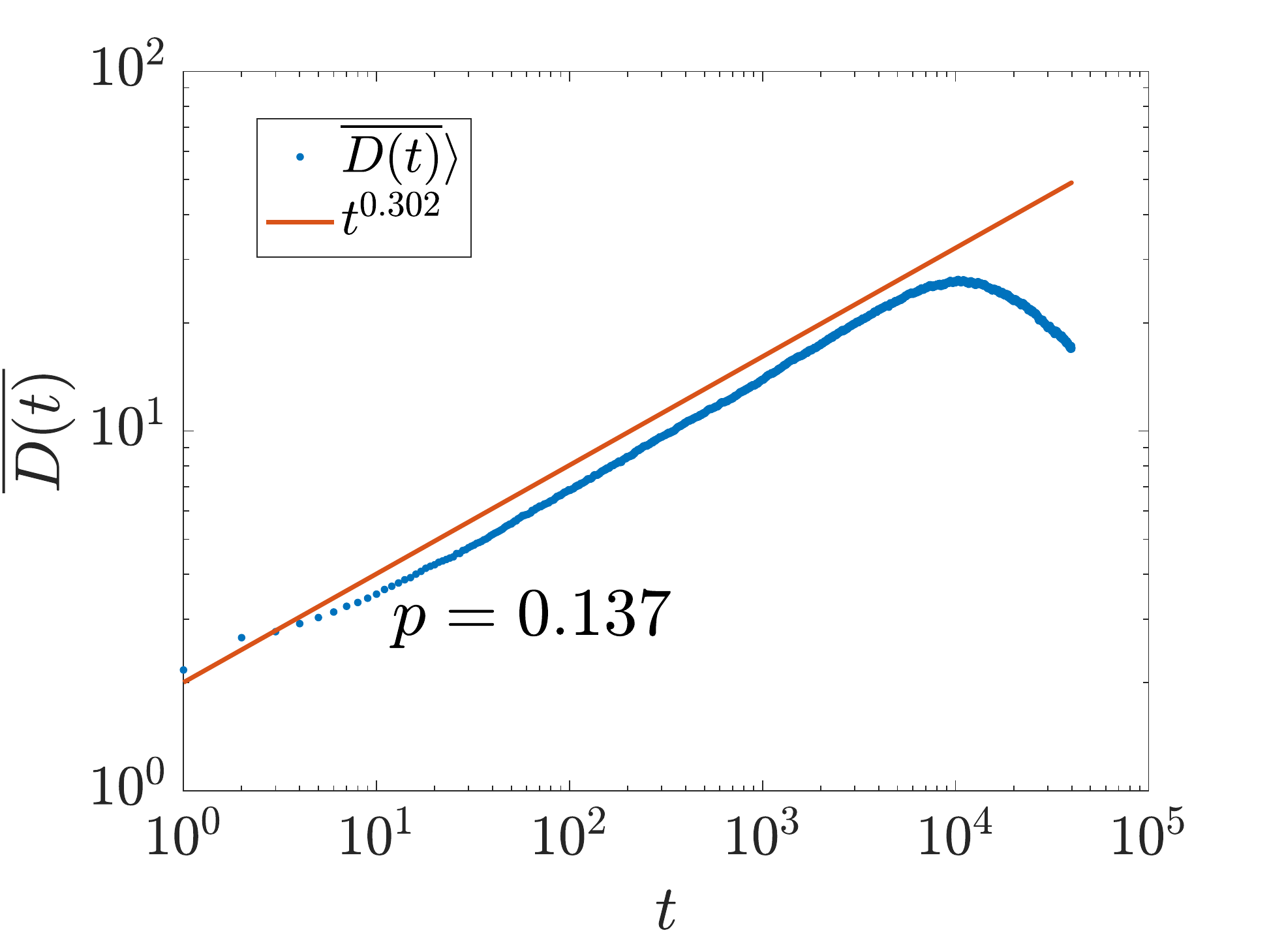}}
\caption{(a) We consider the initial state as two random bit-strings with $\overline{ D(t=0)}\approx L/2$ and compute the averaged Hamming distance $\overline{D(t)}$ as a function of $t$. At the critical point, we find that $\overline{D(t)}$ decays algebraically with exponent close to $0.1595$. In (b)-(d), we consider two bit-strings with the only difference in the middle, i.e., $D(t=0)=4$ and study the dynamics of $\overline{h(x,t)}$ and $\overline{D(t)}$. (b) is the result for $\overline{h(x,t)}$ at $p=0.1<p_c$. It spreads out linearly in time. This is further confirmed by the data collapse in (c). The inset of (b) shows that $\overline{D(t)}$ grows linearly in time and saturates to a constant. (d) At the critical point $p=0.137$, we find that $\overline{D(t)}$ grows as a power law in time and is consistent with the theoretical prediction $t^{0.302}$. We further notice that after long time evolution, $\overline{D(t)}$ will decay in time as in (a). The crossover behavior is known as the critical initial slip. In all these simulations, we take $L=1000$ with open boundary conditions.} 
\label{fig:BS}
\end{figure*}

 We now consider the classical bit-string model which is referenced throughout this work. We demonstrate that this model is in the DP universality class by showing that it has the same critical exponents as the bond DP model. We consider a pair of bit-strings $\ket{m_1}$ and $\ket{m_2}$, which undergo the same stochastic dynamics  described in Fig.~\ref{fig:Clifford_purification_3}. For the bit-string $\ket{m}$, at each site, it takes two possible values: 0 and 1. There are two types of CNOT gates in this classical dynamics: CNOT$_L$ and CNOT$_R$. The CNOT$_L$ gate gives rise to the following update rule:
\begin{align}
    \ket{00}\to \ket{00},\ket{01}\to\ket{01},\ket{10}\to\ket{11},\ket{11}\to\ket{10}.
\end{align}
For CNOT$_R$, it has the following update rule:
\begin{align}
    \ket{00}\to \ket{00},\ket{01}\to\ket{11},\ket{10}\to\ket{10},\ket{11}\to\ket{01}.
\end{align}
We apply CNOT$_L$ and CNOT$_R$ randomly with equal probability. The measurement gate on site $i$ will force $\ket{m_1}$ and $\ket{m_2}$ to take the same value at site $i$. We define the Hamming distance as 
\begin{align}
    D(t)=\sum_{i=1}^L|m_{1,i}-m_{2,i}|.
\end{align}
We compute $\overline{D(t)}$ at different measurement rates $p$ and study its scaling form. Notice that $\overline{D(t)}$ is analogous to $\overline{N(t)}$ in bond DP and can be treated as the order parameter in the dynamical phase transition. 

We first consider the initial condition in which $\ket{m_1}$ and $\ket{m_2}$ are two random bit-strings with $\overline{D(t=0)}=L/2$. We observe a phase transition at $p_c=0.137$. When $p<p_c$, the saturating value $\overline{D(t\to \infty)/L}$ is finite. When $p>p_c$, it decays to zero in a finite amount of time. At the critical point, as shown in Fig.~\ref{fig:BS_purification}, $\overline{D(t)}$ decays algebraically as $D(t)\sim t^{-\beta/\nu_\parallel}$. The power law exponent is $(\beta/\nu_\parallel) = 0.1595$, indicating that this belongs to the DP universality class. Notice that the dynamics starting from this initial condition can be compared with the quantum purification dynamics studied in the main text. 

\begin{figure}[t]
\centering
\subfigure[]{\label{fig:Clifford_v2_purification} \includegraphics[width=.6\columnwidth]{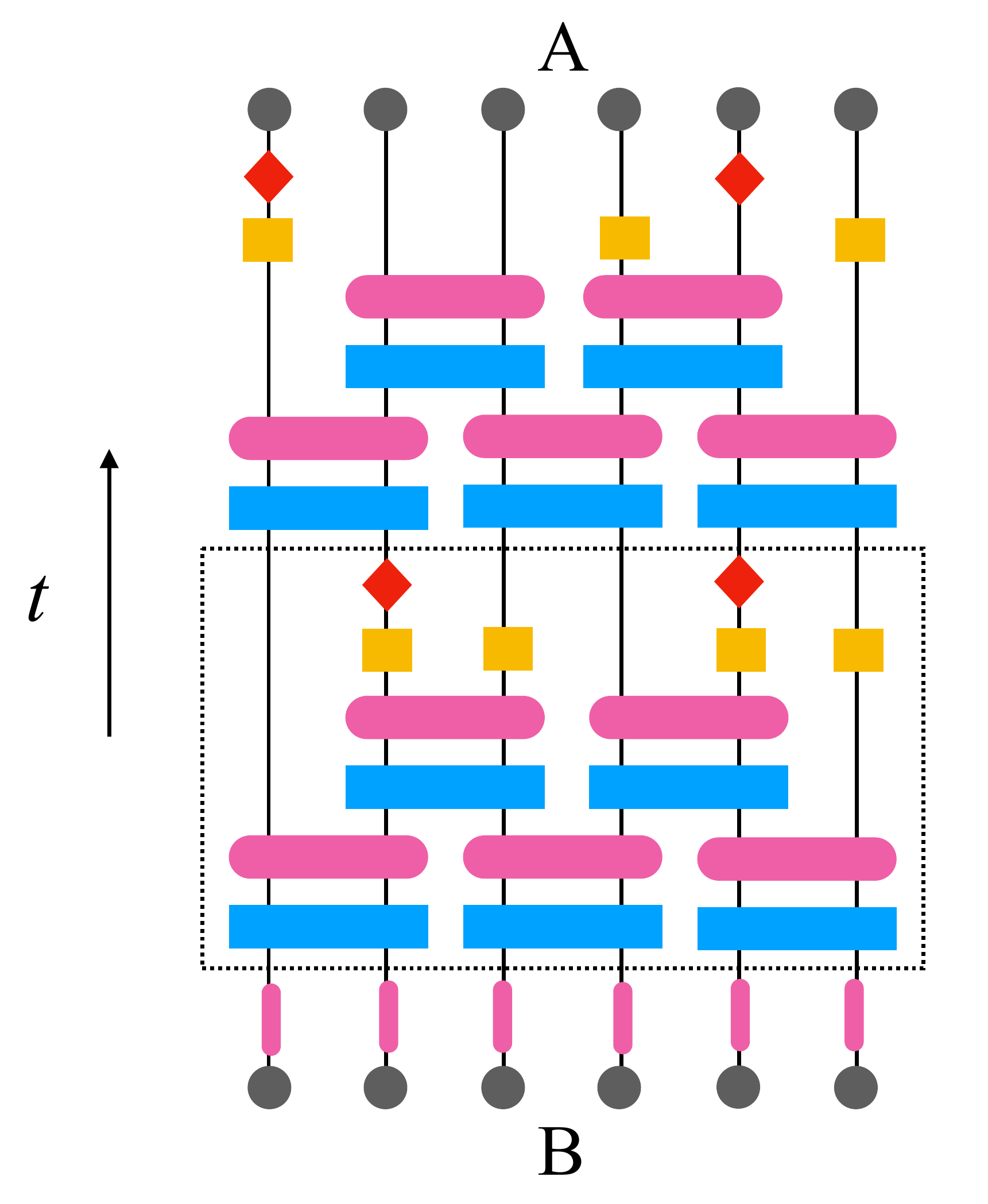}}
\subfigure[]{\label{fig:Clifford_v2_EE} \includegraphics[width=.6\columnwidth]{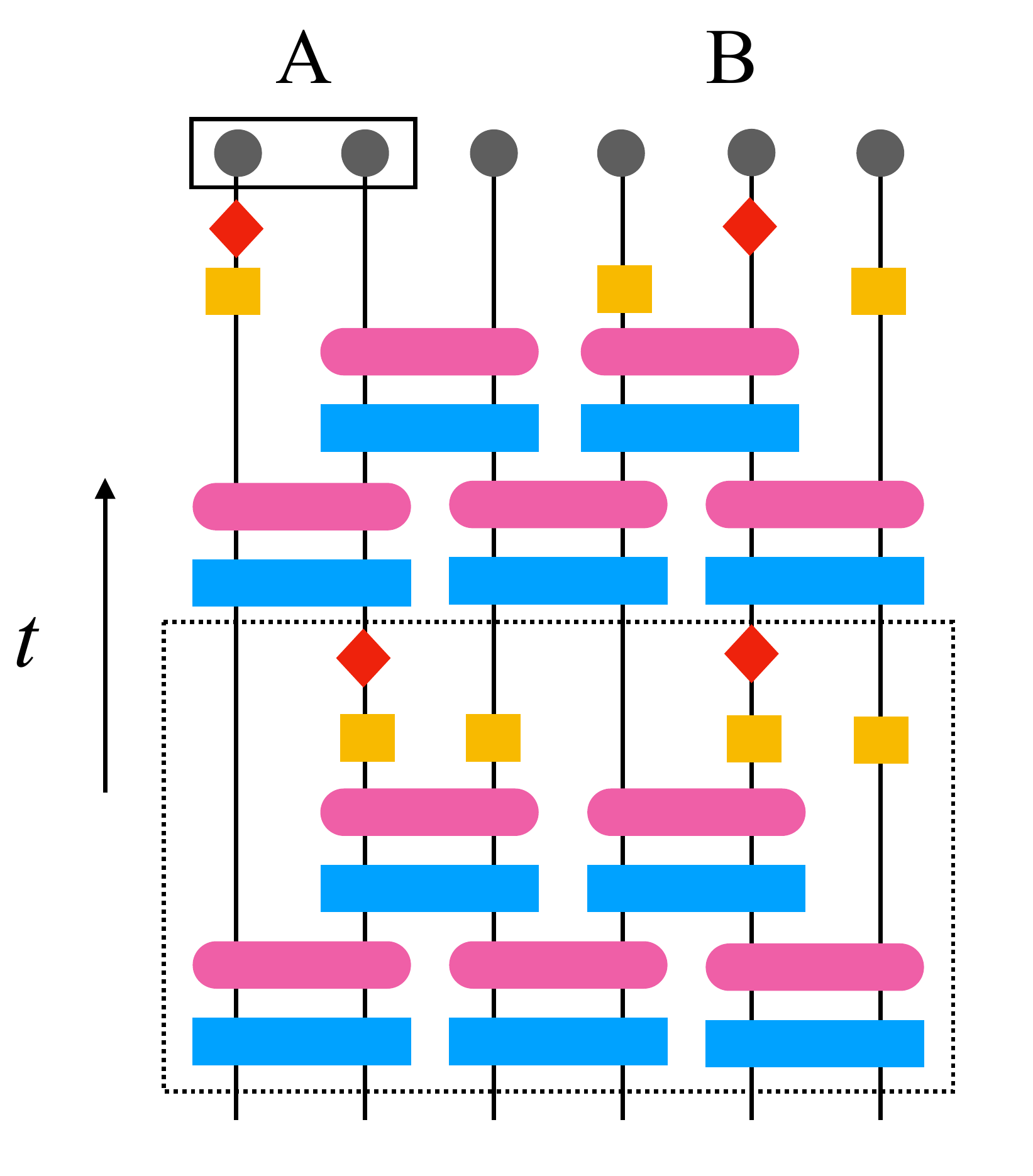}}
\caption{(a) The circuit for the purification dynamics. (b) The circuit for the entanglement dynamics starting from a product state. In both (a) and (b), the dashed box denotes the time evolution in one time step.} 
\label{fig:Clifford_v2}
\end{figure}

We also consider the initial condition in which $\ket{m_1}$ and $\ket{m_2}$ are only differed in the middle. We study how this difference spreads under the time evolution. We characterize this by 
\begin{align}
    h(x,t)=|m_{1,x}-m_{2,x}|.
\end{align}
When $p<p_c$, we find that the front of $\overline{h(x,t)}$ moves linear in time and eventually spreads over the entire system (see Fig.~\ref{fig:BS_spreading}). The data collapse in Fig.~\ref{fig:BS_spreading_collapse} indicates that the front has diffusive broadening. When $p>p_c$, the front  dies out quickly after finite time evolution. At the critical point, we compute 
\begin{equation}
\overline{D(t)}=\sum_x \overline{h(x,t)}
\end{equation} and we find that it increases as a power law in time as $t^{\Theta}$, with $\Theta=0.3$. This is again consistent with the DP universality class critical exponent (See Fig.~\ref{fig:BS_seed}). At late times, it will once again decay algebraically as $t^{-0.1595}$. The motivation of this simulation is to understand the entanglement dynamics starting from a product state studied in the main text.

\section{Clifford circuit with non-QA gate}
\label{sec:non-QA}
\begin{figure*}[hbt]
\centering
\subfigure[]{\label{fig:puri_v2_collapse} \includegraphics[width=.9\columnwidth]{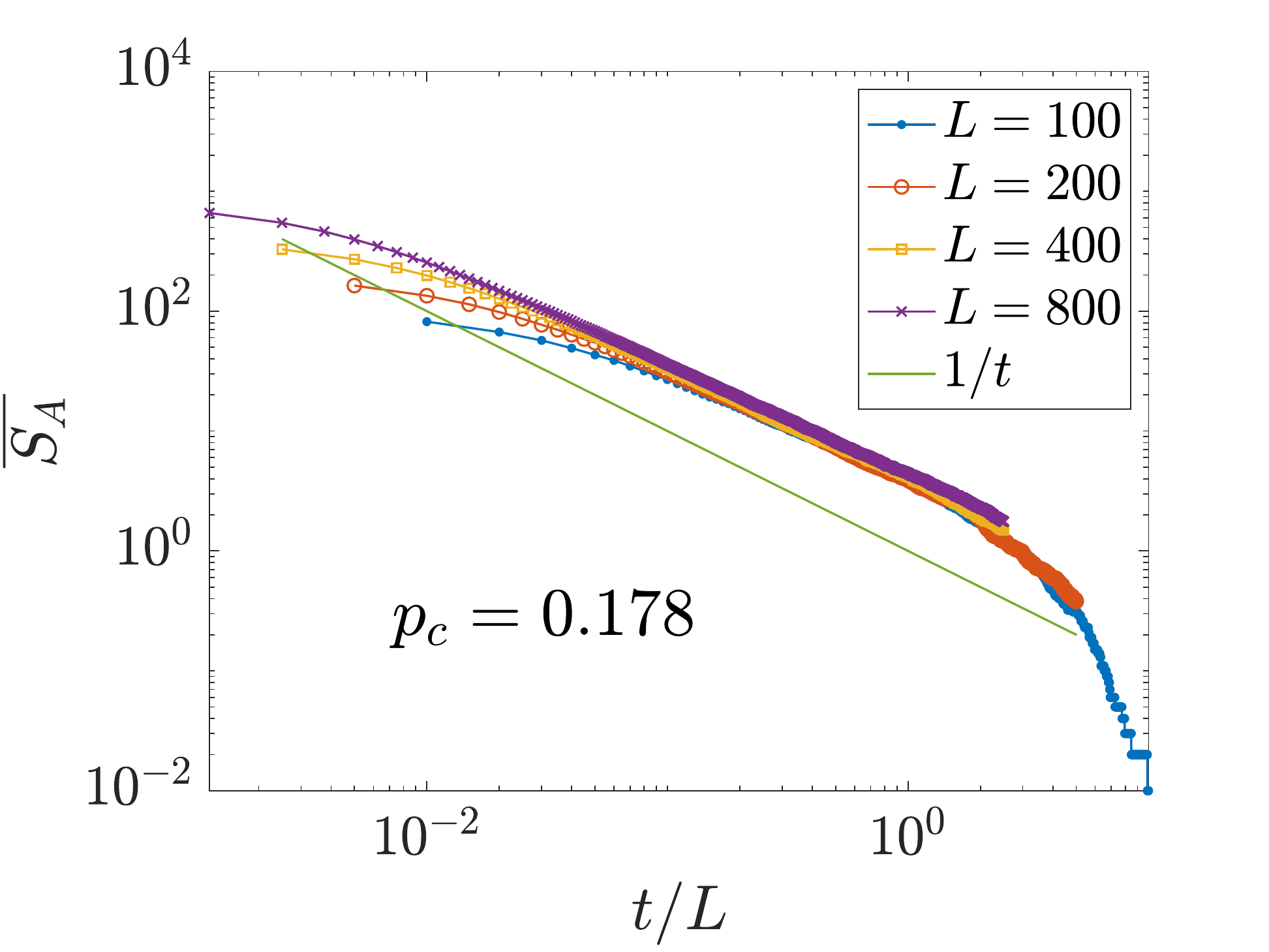}}
\subfigure[]{\label{fig:EE_v2_scaling} \includegraphics[width=.9\columnwidth]{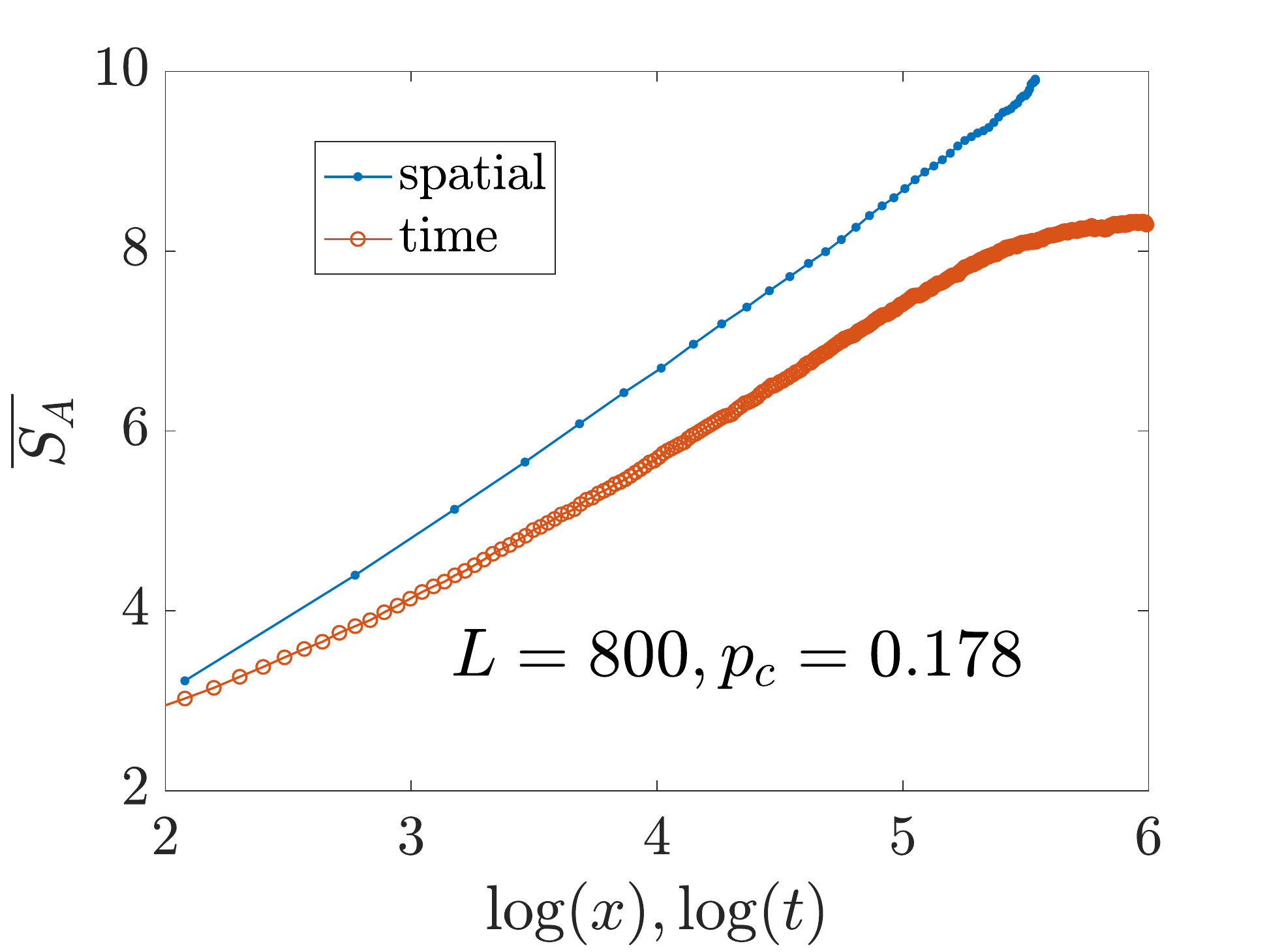}}
\subfigure[]{\label{fig:MI_eta_v2} \includegraphics[width=.9\columnwidth]{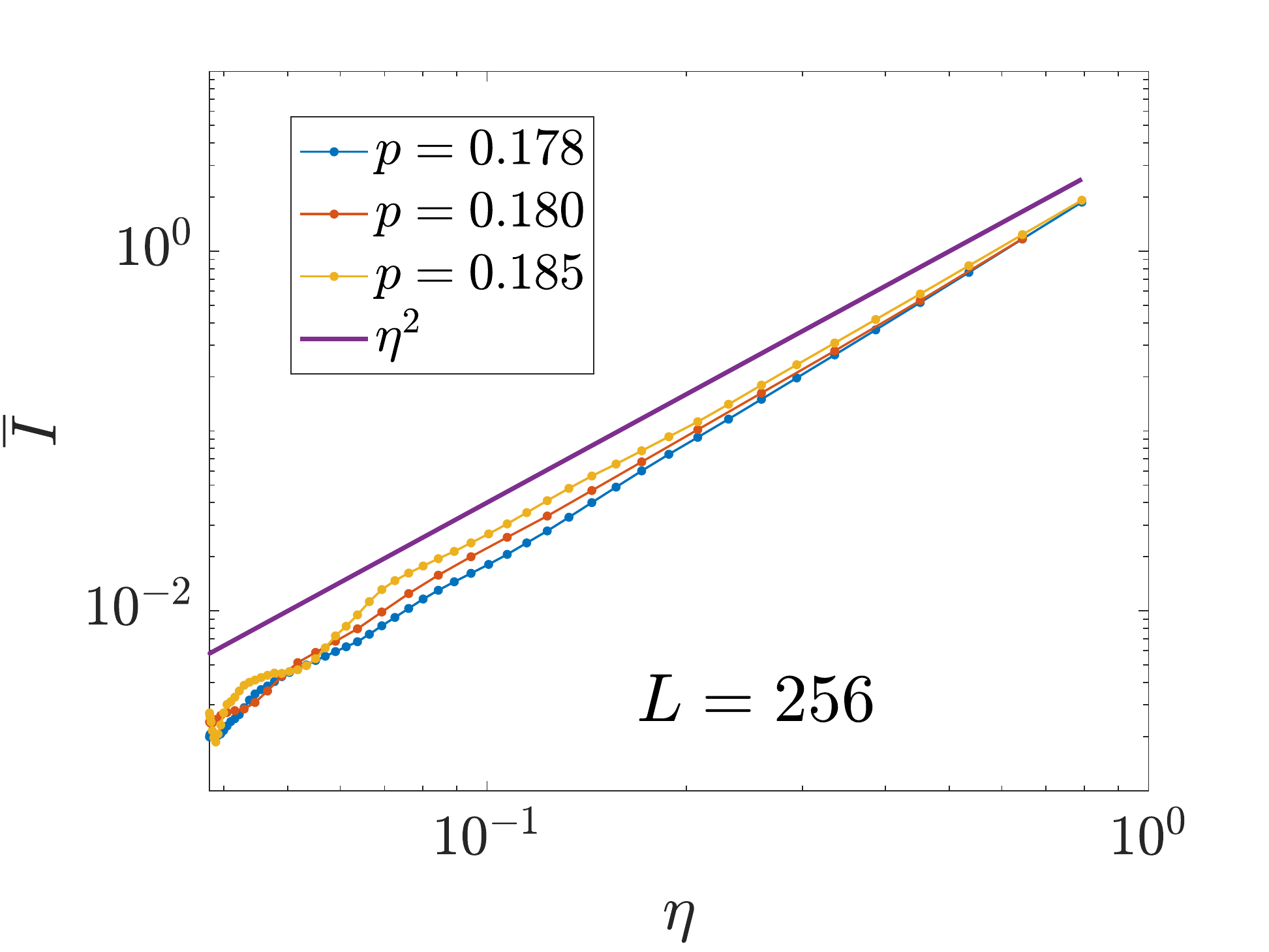}}
\caption{(a) The data collapse of the purification dynamics at the critical point $p=0.178$. (b) The blue curve is the result for the steady state entanglement scaling vs $\log(x)$, where $x\equiv\sin(\pi L_A/L)L/\pi$. The red curve is the result for the early time entanglement entropy vs $\log(t)$. These two curves have the same slope. (c) The mutual information vs cross ratio $\eta$ around the critical point.}
\label{fig:EE_v2}
\end{figure*}

In this appendix, we modify the QA Clifford circuit slightly and introduce a layer of H gates (with $50\%$ probability at each site) in each step of unitary evolution before the Z measurement (See Fig.~\ref{fig:Clifford_v2_EE}). This new Clifford circuit model no longer belongs to the QA class of circuits. We first study the purification dynamics and present the results in Fig.~\ref{fig:puri_v2_collapse}. The data collapse result indicates that the entanglement entropy for the system A has $\overline{S_A(t)}\sim L/t$. The dynamical exponent has $z=1$  We also study the entanglement dynamics from the product state and as shown in Fig.~\ref{fig:EE_v2_scaling}, we have $\overline{S_A(t)}=\alpha_2 \log(t)$. The steady state has entanglement entropy $\overline{S_A}=\alpha_1\log(x)$ with $\alpha_1=\alpha_2$. We further present the mutual information result for the steady state in Fig.~\ref{fig:MI_eta_v2} and we find that $\overline{I}\sim \eta^2$. All these results are the same as for the random Clifford circuit studied in Ref.~\onlinecite{li2020conformal}.

\end{appendix}
\bibliography{QA}
\end{document}